\documentclass[prx,nofootinbib,superscriptaddress,twocolumn,longbibliography]{revtex4-1}

\usepackage{amsmath}
\usepackage{xcolor}
\usepackage{amssymb}
\usepackage{amsxtra,color}
\usepackage{booktabs}
\usepackage{latexsym}
\usepackage{dsfont}
\usepackage[normalem]{ulem} 
\usepackage{array}
\makeatletter
\@ifundefined{selectlanguage}
  {\newcommand{\selectlanguage}[1]{}}
  {\renewcommand{\selectlanguage}[1]{}}
\makeatother

\usepackage{graphicx}
\usepackage{mathtools}
\usepackage{booktabs}

\DeclareUnicodeCharacter{0394}{\ensuremath{\Delta}}
\DeclareUnicodeCharacter{0393}{\ensuremath{\Gamma}}

\graphicspath{ {attoimages/} }   
\usepackage{braket}
\usepackage{tikz}
\usetikzlibrary{arrows}
\usepackage{pgfplots}
\usepgfplotslibrary{groupplots}
\pgfplotsset{compat=1.3}
\usepackage{lipsum}
\usepackage{color}
\usepackage{comment}

\usepackage{float}

\usepackage[T1]{fontenc}
\usepackage{lmodern}
\usepackage[utf8]{inputenc} 
\usepackage{MnSymbol}	

\usepackage[unicode]{hyperref}

\definecolor{MyDarkGreen}{rgb}{0,0.6,0}
\definecolor{MyDarkBlue}{rgb}{0,0,0.8}
\definecolor{MyDarkRed}{rgb}{0.6,0,0.3}

\hypersetup{breaklinks=true, colorlinks=true,plainpages=true, linktocpage=true, linkcolor=MyDarkBlue, citecolor=MyDarkGreen, urlcolor=MyDarkRed, pdfborder={0 0 0},%
pdfauthor={},%
pdfsubject={Research article},
pdftitle={}%
}

\begin{document}

\title{From interface-limited to Auger-dominated carrier dynamics in $\pi$-SnS}
\author{Hugo \surname{Laurell}$^{\dag}$}
\email{hugo.laurell@fysik.lu.se}
\affiliation{Department of Chemistry, University of California, Berkeley, California, 94720, USA}
\affiliation{Department of Physics, University of California, Berkeley, California, 94720, USA}
\affiliation{Material Sciences Division, Lawrence Berkeley National Laboratory, Berkeley, California 94720, USA}
\affiliation{Department of Physics, Lund University, Box 118, 22100 Lund, Sweden}
\author{Kevin \surname{Xiong}$^{\dag}$}
\affiliation{Department of Chemistry, University of California, Berkeley, California, 94720, USA}
\author{Nedjma \surname{Ouahioune}$^{\dag}$}
\affiliation{Department of Physics, Lund University, Box 118, 22100 Lund, Sweden}
\author{Thomas \surname{Kjellberg Jensen}$^{\dag}$}
\affiliation{Department of Physics, Lund University, Box 118, 22100 Lund, Sweden}
\author{Jonah R. \surname{Adelman}}
\affiliation{Department of Chemistry, University of California, Berkeley, California, 94720, USA}
\author{Kylie J. \surname{Gannan}}
\affiliation{Department of Chemistry, University of California, Berkeley, California, 94720, USA}
\author{Rafael \surname{Quintero-Bermudez}}
\affiliation{Material Sciences Division, Lawrence Berkeley National Laboratory, Berkeley, California 94720, USA}
\affiliation{Department of Chemistry, University of California, Berkeley, California, 94720, USA}
\author{Lior \surname{Verbitsky}}
\affiliation{Department of Chemistry, University of California, Berkeley, California, 94720, USA}
\author{Han K. D. \surname{Le}}
\affiliation{Department of Chemistry, University of California, Berkeley, California, 94720, USA}
\author{Anders \surname{Mikkelsen}}
\affiliation{Department of Physics, Lund University, Box 118, 22100 Lund, Sweden}
\author{Peidong \surname{Yang}}
\affiliation{Department of Chemistry, University of California, Berkeley, California, 94720, USA}
\affiliation{Material Sciences Division, Lawrence Berkeley National Laboratory, Berkeley, California 94720, USA}
\affiliation{Department of Materials Science and Engineering, University of California, Berkeley, California 94720, USA}
\affiliation{Kavli Energy Nano Science Institute, Berkeley, California 94720, USA}
\author{Carl \surname{Hägglund}}
\affiliation{Department of Materials Science and Engineering, Solar Cell Technology, Uppsala University, Box 35,751 03 Uppsala, Sweden}
\author{Stephen R. \surname{Leone}}
\affiliation{Department of Chemistry, University of California, Berkeley, California, 94720, USA}
\affiliation{Department of Physics, University of California, Berkeley, California, 94720, USA}
\affiliation{Chemical Sciences Division, Lawrence Berkeley National Laboratory, Berkeley, California 94720, USA}

\begin{abstract}
\begin{centering}
Metastable cubic tin(II) sulfide ($\pi$-SnS) is an earth-abundant semiconductor whose three-dimensionally bonded, chiral lattice provides a possible route to mitigate the short minority carrier lifetime in orthorhombic SnS thin films, while it has a bandgap with a sharp onset nearly ideal for the top cell in a tandem device. The high surface-to-volume ratio of thin-film SnS makes carrier lifetimes and many-body relaxation pathways central to device-relevant performance, yet the microscopic mechanisms governing ultrafast cooling and recombination over illumination density remain poorly constrained. Here we use core-level extreme-ultraviolet (XUV) attosecond transient absorption spectroscopy across the Sn $4d$ edge to track carrier injection, cooling, and recombination in $\pi$-SnS with element- and orbital-specific sensitivity. After femtosecond near-infrared excitation, the Sn $4d\rightarrow$CB onset exhibits pronounced conduction-band state filling and a carrier-induced edge shift, enabling extraction of density-dependent kinetics. The transient response is well described by a biexponential decay, revealing a fast component associated with hot-carrier cooling and a slow component associated with recombination. At low carrier densities, the recombination dynamics are consistent with interface-limited processes, whereas above $\sim 1\times10^{20}$~cm$^{-3}$ both cooling and recombination accelerate, indicating a crossover to carrier--carrier interaction-dominated dynamics. In addition, coherent phonon oscillations with a period of $\sim 188$~fs are observed, suggesting coupling between electronic excitation and lattice motion. These results provide a comprehensive picture of nonequilibrium carrier and phonon dynamics in cubic SnS, reveal a change of mechanisms over a range of carrier densities, and establish the value of using attosecond transient absorption spectroscopy to study ultrafast processes in complex semiconductors that have optoelectronic and energy-conversion applications.
\end{centering}
 \end{abstract}
\maketitle

\begingroup
\renewcommand\thefootnote{\dag}
\footnotetext{These authors contributed equally to this work.}
\endgroup

\section{Introduction}
Tin(II) sulfide (SnS) is an earth-abundant semiconductor with a favorable band gap and strong visible-light absorption, making it ideal as the top cell in a Tandem device. Its photovoltaic efficiency in thin films is limited by short minority carrier lifetimes, originating from rapid nonradiative recombination losses at defects and interfaces \cite{Jaramillo2016}.
Extending the minority carrier lifetimes is a route toward improving SnS photovoltaic performance. In its equilibrium form, SnS crystallizes in the layered orthorhombic Pnma structure, which manifests an anisotropic electronic and optical response that can further complicate carrier transport and extraction in thin-film devices \cite{Sine2011,Banai2014}. Recently, a metastable cubic polymorph of SnS, denoted $\pi$-SnS and adopting the chiral space group $P2_{1}3$, was identified and structurally resolved \cite{Abutbul2016_2,Skelton2017_2}. Since then, $\pi$-SnS growth has been demonstrated by a variety of methods, including Chemical Bath Deposition (CBD) \cite{Abutbul2020}, Chemical Spray Pyrolysis (CSP) \cite{Polivtseva2017}, Aerosol-Assisted Chemical Vapor Deposition (AA-CVD) \cite{Ahmet2019} and Atomic Layer Deposition (ALD) \cite{Bilousov2017}. First-principles calculations showed that $\pi$-SnS is dynamically stable and energetically competitive, lying only a few kJ\,mol$^{-1}$ above the orthorhombic ground state, in contrast to dynamically unstable zincblende and rocksalt variants. Importantly, the three-dimensionally bonded cubic lattice of $\pi$-SnS reduces structural anisotropy and provides a more isotropic framework for carrier transport and energy relaxation, potentially mitigating symmetry-driven limitations on carrier cooling and recombination under high excitation densities. Femtosecond transient-absorption studies on orthorhombic SnS have reported biexponential hot-carrier relaxation attributed to coupling to distinct optical-phonon branches, highlighting the rich carrier-cooling pathways accessible in this material family \cite{Peng2025}.

The measurement of ultrafast carrier dynamics in semiconducting materials on attosecond timescales has enabled direct insight into the fundamental interactions governing their optical and electronic properties \cite{FerrayJPB1988,McPhersonJOSAB1987,Palo2024}. In particular, attosecond transient absorption spectroscopy (ATAS) provides simultaneous access to electronic excitation, energy relaxation, and lattice-driven dynamics with elemental and band-specific sensitivity, via high spectral and temporal resolution. Understanding how these processes unfold on their intrinsic timescales is essential for controlling material functionality in optoelectronic and energy-conversion applications. By combining a femtosecond infrared (IR) pump with an attosecond extreme ultraviolet (XUV) probe, ATAS can track a hierarchy of timescales ranging from carrier injection on few-femtosecond timescales, to carrier cooling on $\sim$10$^2$~fs timescales, and ultimately recombination dynamics on picosecond to tens-of-picoseconds timescales. Recent advances in attosecond transient absorption spectroscopy have enabled direct access to coupled electronic and lattice dynamics in solids \cite{Zuerch2017,Cushing2018,Attar2020}. Coherent phonon oscillations provide a sensitive probe of electron–phonon coupling \cite{Adelman2025}, while hot-carrier cooling dynamics govern how excess electronic energy is dissipated on ultrafast timescales \cite{Fu2017,Yang2016,Laurell2025}. Both processes play a central role in governing energy-loss pathways and strongly influence charge transport properties in semiconductors \cite{Alberding2016}.

To date, attosecond studies of solids have often focused on elemental materials or systems with small primitive unit cells, where the ultrafast response can be described in terms of a limited set of electronic bands and vibrational modes \cite{Drescher2025,Geneaux2021}. In contrast, large unit cells result in a dense manifold of coupled electronic and lattice degrees of freedom. Only recently has attosecond spectroscopy begun to address such complex solids, for example multinary semiconductors with complex defect structures \cite{Laurell2025}. Extending attosecond methodologies to this regime therefore represents an important frontier in solid-state attosecond science.

In this work, we use attosecond transient absorption extreme ultraviolet (XUV) spectroscopy to investigate ultrafast carrier and lattice dynamics in $\pi$-SnS. Coherent phonon oscillations are observed following excitation across the optical bandgap with a period of 188~fs, revealing coupling between electronic excitation and lattice motion. In addition, biexponential decays are observed in the transient absorption signal, which we attribute to cooling and interband recombination of carriers. In particular, the investigations emphasize measurements as a function of carrier density, revealing a change of mechanism from interface- to Auger dominated with increasing carrier density. Together, these results provide new insights into the nonequilibrium dynamics of $\pi$-SnS and establish this material as a compelling model system for studying ultrafast processes in emerging semiconductors.

\begin{figure*}[htbp!]
    \centering
    \includegraphics[width=1\linewidth]{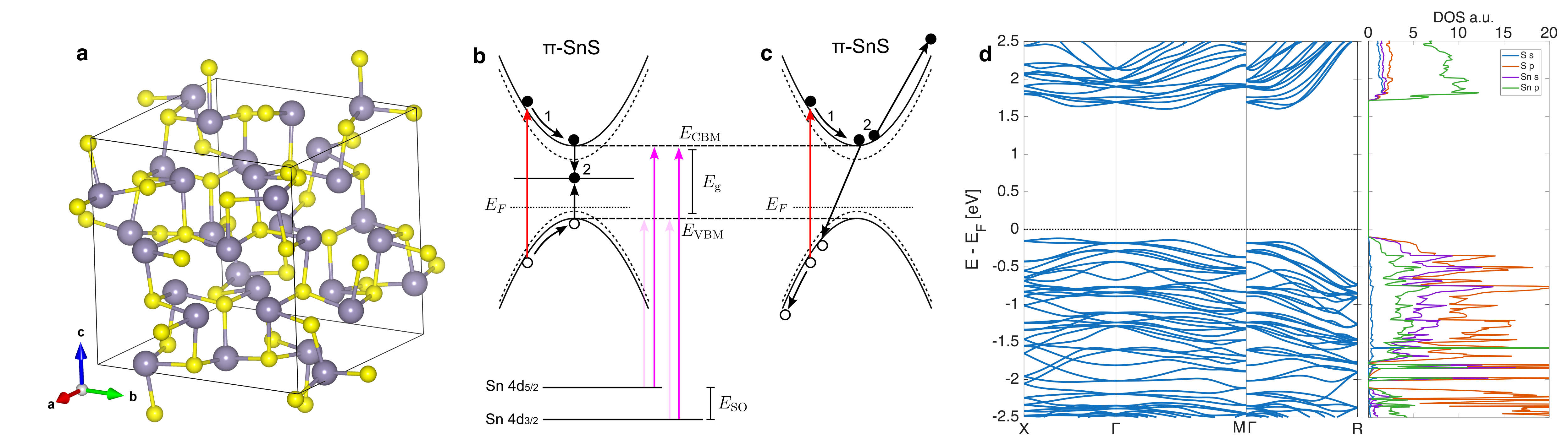}
    \caption{Crystal structure, excitation pathways, and electronic structure of cubic $\pi$-SnS. (a) Real-space crystal structure of cubic $\pi$-SnS, illustrating the three-dimensional network consisting of 64 Sn (grey) and S (yellow) atoms within the P3$_1$2 unit cell. (b) Schematic energy-level diagram summarizing the photoexcitation and relaxation processes probed in this work. Near-infrared excitation across the indirect band gap (red arrow) generates hot electron and hole populations in the conduction and valence bands, respectively. These carriers undergo rapid intraband cooling toward the band edges (1), followed by recombination mediated by interface or defect states within the band gap (discrete level), and subsequent electron–hole recombination (process 2). (c) Energy diagram for Auger cooling (1) and Auger recombination (2). The dashed parabolas in (b) and (d) show the shift of the bands induced by bandgap renormalization. (d) Density-functional-theory (DFT) calculated electronic band structure of cubic $\pi$-SnS along high-symmetry directions of the Brillouin zone, together with the projected density of states (DOS), highlighting the band gap and the orbital character of the valence and conduction bands relevant to the XUV transient absorption measurements. The crystal structure (a) and DFT calculated bandstructure (d) was adapted from \cite{Skelton2017_2}.}
    \label{fig:fig_1}
\end{figure*}

Fig.~\ref{fig:fig_1} summarizes the structural and electronic framework underlying the core-level XUV transient absorption measurements presented in this work. Fig.~\ref{fig:fig_1}(a) shows the real-space crystal structure of cubic $\pi$-SnS, which forms a three-dimensionally bonded network in contrast to the layered orthorhombic (Pnma) phase and crystallizes in the chiral, non-centrosymmetric space group $P2_{1}3$ \cite{Skelton2017}. The schematic in Fig.~\ref{fig:fig_1}(b) illustrates the photoexcitation and relaxation pathways probed in the pump--probe experiment. Near-infrared excitation across the band gap (red arrow) generates hot electron and hole populations, which undergo rapid intraband relaxation toward the band edges (process~1). On longer timescales, the conduction-band population is depleted by recombination, which in thin films can be mediated by interface or defect states within the gap (process~2) and, at sufficiently high photoexcited carrier densities, can be supplemented by multicarrier (Auger) pathways. Fig.~\ref{fig:fig_1}(c) schematically depicts these multicarrier channels, distinguishing Auger-assisted intraband energy redistribution (Auger cooling, process~1) from carrier annihilation via Auger recombination (process~2). Finally, Fig.~\ref{fig:fig_1}(d) presents the DFT-calculated band structure and projected density of states of $\pi$-SnS. These calculations form the basis for assigning the observed XUV spectral features to Sn $4d \rightarrow$ conduction-band minumum (CBM) transitions and for interpreting the transient absorption signal as a phenomenological, element- and orbital-specific measure of conduction-band state filling and its subsequent decay.

\section{Methods}

\subsection{Sample synthesis}

SnS was grown on the front side of Si$_3$N$_4$ TEM windows (Norcada), using a viscous flow, hot wall reactor (Picosun R-200 Advanced ALD system). Growth on the back of the windows was suppressed by supporting the chips on a clean silicon wafer, resulting in very tight gap between the two near atomically smooth surfaces. Tin(II) acetylacetonate [Sigma-Aldrich 99.9\%] and hydrogen sulfide (Air Liquide, 99.5\%) were used as precursors, with nitrogen (N$_2$, 99.9999\%) as the carrier gas. The substrate and reactor chamber were maintained at 120 °C, and the Sn(acac)$_2$ source at 100 °C. 

In order to drive the SnS film growth towards the cubic phase, 2 ALD cycles of trimethylaluminum (TMA, electronic grade, Pegasus Chemicals) and deionized water were performed to surface modify the substrate with an Al$_2$O$_3$-like layer. Between 750 and 1200 cycles of SnS ALD were subsequently performed without breaking vacuum, resulting in $\pi$-SnS films with thicknesses from 32 to 45 nm, as determined by spectroscopic ellipsometry performed on the parallel processed Si substrate. The ellipsometry analysis also confirmed a bandgap and absorption onset behavior characteristic of the $\pi$-SnS polymorph. Further details on the ellipsometry modeling and the stop-flow sequence used for the SnS ALD cycle are provided in prior publications \cite{Bilousov2021,Bilousov2017}.

\subsection{Core-level XUV transient absorption measurements}

A Ti:Sapphire laser system (Coherent Astrella) outputs 35 fs, 3.0 mJ pulses with a central wavelength of 800 nm at 1 kHz repetition rate. These pulses are sent through a stretched 2.2 m long hollow core fiber filled with argon gradient (90 mTorr at entrance, 6 Torr at exit) for spectral broadening through self-phase modulation. Subsequent chirped mirrors (8 pairs) and a pair of potassium dihydrogen phosphate wedges compensate for $2^{nd}$ and $3^{rd}$ order dispersion, respectively, and produced 4 fs, 500-900 nm near-infrared (NIR) pulses. This serves as both the pump arm and the driving field for high-order harmonic generation (HHG) in the probe arm, which are split by a 10:90 beam-splitter. In the probe arm, the NIR pulse is focused into a krypton filled gas cell (23 Torr) with a $f = 45$ cm focal length mirror and generated few-burst attosecond pulses through HHG. The driving NIR is filtered out with a 70 nm thick aluminum foil. Then, the XUV probe arm is refocused by a toroidal mirror and spatially recombined at the SnS sample with NIR pump arm, latter of which is focused by a 1 m focal length mirror. Behind the samples, a secondary 70 nm thick aluminum foil filters out residual NIR pump and scattered light. Any transmitted XUV is spatially dispersed through diffraction off a flat-field grating (01-0639, Hitachi) onto a charge-coupled device camera (PIXIS 400 B). SnS samples are first photoexcited by the NIR pump and then probed at a time delay $\tau$ (controlled by a retroreflector on a translation stage) by recording the transmission of the XUV probe. Measurements are performed for pump energies ranging from 3 $\mu$J to 11 $\mu$J, corresponding to charge carrier densities from $1.6\times10^{19}$~cm$^{-3}$ to $3.1\times10^{20}$~cm$^{-3}$ assuming linear absorption.

To minimize laser heating and avoid photoinduced phase transformation \cite{Hegde2020,Kishore2023}, as verified by x-ray diffraction (XRD) and Raman, discussed further below, the laser pulses are mechanically chopped to an effective repetition rate of 100~Hz. Previous measurements performed at the 1~kHz repetition rate resulted in irreversible structural changes of the SnS films, characterized by a transition from the metastable cubic $\pi$-SnS phase to the thermodynamically stable orthorhombic phase. Operating at a reduced effective repetition rate therefore ensures that the deposited heat load remains below the threshold for phase transformation during the attosecond transient absorption measurements.

To verify the structural integrity of the samples under these experimental conditions, XRD and Raman spectroscopy are performed after completion of the attosecond transient absorption measurements. The post-measurement XRD patterns and Raman spectra, presented in the SM section \ref{raman} and \ref {xrd}, confirm that the samples retained the cubic $\pi$-SnS phase and showed no signatures of conversion to the orthorhombic polymorph. These measurements demonstrate that the reduced repetition-rate excitation protocol preserves the metastable crystal structure throughout the experiment.

\section{Results and discussion}

In this section, we present attosecond transient absorption measurements of photoexcited $\pi$-SnS, probing carrier and lattice dynamics via Sn $4d$ core-to-band transitions. The broadband XUV probe provides energy-resolved access to conduction-band state filling, carrier cooling, and recombination on femtosecond to picosecond timescales. We analyze the density-dependent carrier relaxation dynamics and show that the transient signal also reveals coherent phonon motion, allowing direct extraction of phonon frequency, phase and dephasing time.

\begin{figure}[htbp!]
    \includegraphics[width=0.9\linewidth]{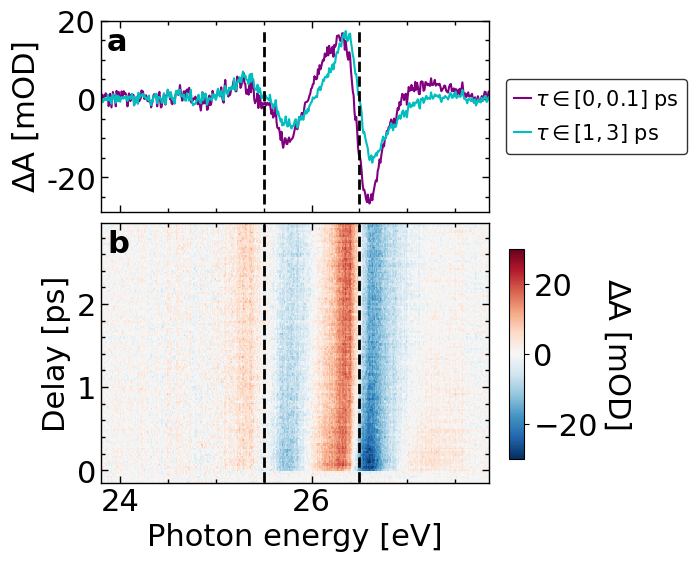}
    \caption{(a) Differential XUV absorbance of $\pi$-SnS across the Sn N$_{4,5}$ absorption edge. The differential absorbance is shown as function of XUV photon energy averaged over pump-probe delays between 0 and 100 fs (purple curve) and between 1 and 3 ps (blue curve). (b) Differential XUV absorbance as a function of the photon energy and pump-probe delay. The dashed vertical lines indicate the energies of the 4d$_{5/2}$$\rightarrow$CBM and 4d$_{3/2}$$\rightarrow$CBM transitions.}\label{fig:fig_2}
\end{figure}

In Fig.~\ref{fig:fig_2}(b) an example is displayed of an attosecond transient absorption scan of $\pi$-SnS at a peak charge carrier density of $7.1\times10^{19}$~cm$^{-3}$ (carrier density calculation using Eq. \eqref{SM:eq:ccdens} in the SM), showing the differential XUV absorbance ($\Delta A(\epsilon,\tau)$) as a function of pump--probe delay and XUV photon energy. Negative delays, corresponding to the XUV probe arriving before the IR pump, show no observable differential absorption after subtraction of the thermal background. At temporal overlap between the IR pump and XUV probe ($t=0$), strong variations in the XUV absorbance appear and persist for several picoseconds, well beyond the pump--probe overlap.

The largest variations in $\Delta A(\epsilon,\tau)$ occur at photon energies of 25.5 eV and 26.5 eV, corresponding to the Sn 4d$_{5/2}$$\rightarrow$CBM and 4d$_{3/2}$$\rightarrow$CBM core-level transitions (black dashed lines) \cite{Padova1994,Wertheim1989,Taniguchi1990}. Importantly, these two spin-orbit split transitions are to a large extent separated in energy and exhibit minimal spectral congestion, enabling the transient dynamics associated with each core-to-band transition to be resolved independently. This splitting agrees well with the 1.056~eV value previously measured by x-ray photoelectron spectroscopy \cite{Wertheim1989}.

At energies near the two core-level transitions, pronounced changes in $\Delta A(\epsilon.\tau)$ are observed. For both spin--orbit components, the XUV absorbance decreases (increases) at photon energies slightly above (below) the absorption edge. We attribute this behavior to a combination of band-gap renormalization, pump-induced shifts of the core-level potential energy driven by lattice motion, and state filling and opening of conduction-and valence band states respectively \cite{Zuerch2017}, as illustrated schematically in Fig.~\ref{fig:fig_1}(b). Figure~\ref{fig:fig_2}(a) further shows the differential absorbance averaged over short (0--100~fs) and long (1-3~ps) delays. The reduced magnitude of $\Delta A(\epsilon,\tau)$ at later times reflects electron--hole recombination and lattice relaxation.

The bandgap of $\pi$-SnS was determined to be $E_g=$1.67 eV through ellipsometry measurements. Therefore, we expect features corresponding to XUV transitions to the valence band (VB) maximum from the 4$d$ core shell at 24.8 eV and 23.8 eV, as illustrated by the lighter pink arrows in Fig. \ref{fig:fig_1}(b). However, we do not observe signal at these energies. The absence of a detectable hole-related signal in the XUV transient absorption scan can be understood from both dipole-selection rules and the orbital character of the electronic bands in $\pi$-SnS. As shown in Fig.~\ref{fig:fig_1}(d), the VB is composed predominantly of a mixture of Sn-$5s$ and S-$3p$ orbitals, with a lesser contribution of Sn-$4p$ orbitals \cite{Skelton2017_2}, whereas the Sn $4d$ core level only allows dipole-allowed transitions to $p$-like final states for linearly polarized XUV light. Transitions from Sn $4d$ to the predominantly Sn $s$-type VB states are dipole forbidden, possibly explaining the absence of a pronounced transient absorption signature associated with photoinduced holes in the largely $5s$ valence band. In contrast, the CBM exhibits strong Sn-$5p$ character, as evident from the projected density of states in Fig.~\ref{fig:fig_1}(d), making the Sn $4d \rightarrow$ CBM transitions both dipole allowed and highly sensitive to changes in the conduction-band electron population. As a result, the Sn edge provides a direct, element- and orbital-specific probe of conduction-band state filling, intraband relaxation, and recombination dynamics.

From a photovoltaic perspective, this selectivity is particularly relevant, since carrier transport, extraction, and recombination losses are predominantly governed by the dynamics of electrons near the conduction-band minimum. The Sn $4d$-edge transient absorption, therefore, can directly reflect electronic processes that are central to photovoltage generation and carrier collection, rather than providing an indirect optical proxy. Combined with the ultrafast temporal resolution of the attosecond probe, this makes the Sn edge an ideal observable for tracking photophysically and technologically relevant carrier dynamics in $\pi$-SnS.

\subsection{Ultrafast build-up of the conduction-band response}

\begin{figure}[htbp!]
    \centering
    \includegraphics[width=1\linewidth]{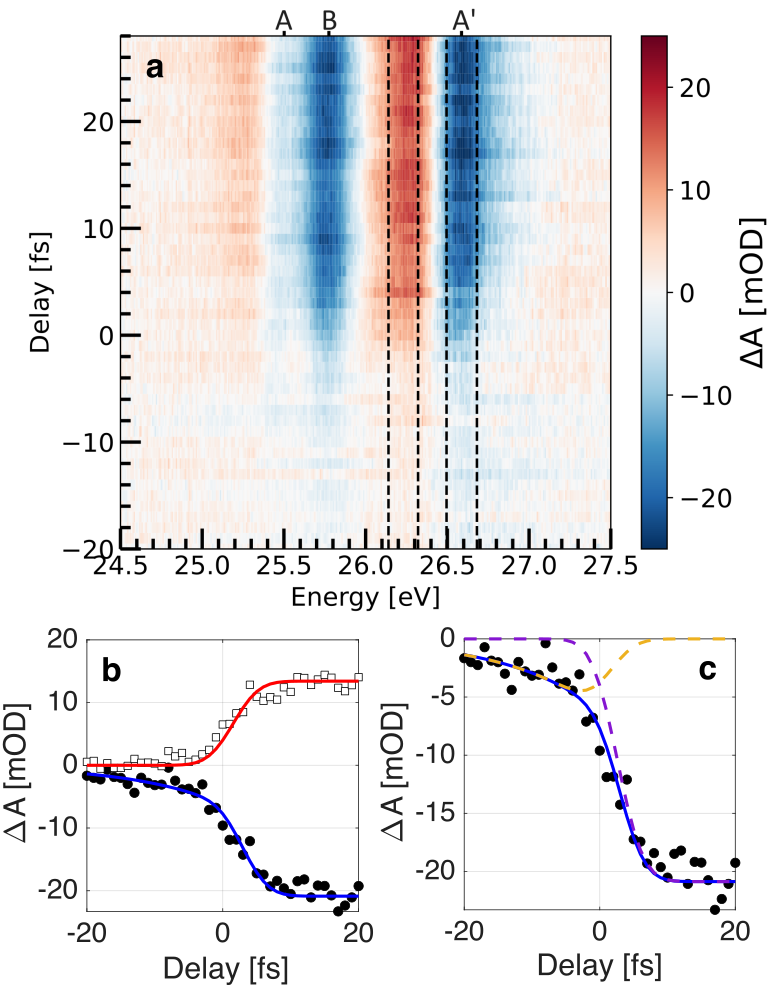}
    \caption{Ultrafast build-up of the Sn $4d\rightarrow$CB onset response in $\pi$-SnS. (a) Short-timescale ATAS scan across the Sn $4d$ absorption onset recorded with 1~fs delay steps with an exciting pump energy of 9 $\mu $J. The indicated spectral windows highlight a state-filling window (dashed region at A$'$) and an edge-shift window (left dashed region). Both windows include contributions from pump-induced renormalization of the Sn $4d\rightarrow$CB onset (including band-gap and core-level shifts and core-exciton renormalization), while the green window additionally tracks conduction-band state filling. (b) Corresponding temporal responses illustrating an approximately instrument-response-limited rise of the state filling signal and the build-up of the edge-shift response, as well as a decay towards negative times for the state blocking signal. (c) Decomposition of state filling signal. The fitted A$'$ core-exciton and carrier injection response is shown as dashed gold and dashed purple curves respectively.}
    \label{fig:SM_shortScan}
\end{figure}

The rise of the differential absorption response across the Sn $4d\rightarrow$CBM transition onset is governed by a combination of conduction-band state filling and an electronic red shift of the absorption edge driven by carrier-induced band-gap renormalization (BGR) (shown as dashed parabolas in Fig. \ref{fig:fig_1}(b,c)). To resolve how these contributions manifest in the transient absorption observable, Fig.~\ref{fig:SM_shortScan}(a) shows a short-delay ATAS scan recorded with a 1~fs delay step around the Sn $4d\rightarrow$CB onset at a carrier density of $1.9\times10^{20}$~cm$^{-3}$. Immediately following NIR excitation, photoexcited electrons populate the Sn-$5p$ conduction-band states, probed by the XUV Sn $4d\rightarrow$CB transitions, resulting in a negative $\Delta A(\epsilon,\tau)$ associated with state filling and BGR (A$'$ region in Fig. \ref{fig:SM_shortScan}(a)). The resulting nonequilibrium electron--hole distributions give rise to strong carrier--carrier interactions that renormalize the electronic band structure on an ultrafast timescale, leading to a reduction of the band-gap energy and a red shift of the absorption onset (positive $\Delta A(\epsilon,\tau)$ left to the A$'$ region Fig. \ref{fig:SM_shortScan}(a))~\cite{Schultze2014}. As the carrier density subsequently decreases through recombination, the screening weakens, and the red-shifted absorption edge relaxes back toward its equilibrium position. 

Furthermore, since the Sn N$_{4,5}$ absorption edge supports core-excitons (A, B and A$'$), the transient edge response includes a strong modification of the lineshape arising from carrier-induced renormalization of the core-exciton resonances. Photoexcited carriers alter the screened Coulomb interaction that binds the Sn N$_{4,5}$ core exciton, leading to pump-induced shifts and distortions of the core-to-band transition that add to $\Delta A(\epsilon,\tau)$. Recent core-level transient absorption studies have shown that Coulomb screening by photoexcited carriers plays a dominant role in shaping core-exciton dynamics across a wide range of materials and excitation conditions, with Pauli blocking via state filling acting as a secondary contribution~\cite{Rossi2025}. Accordingly, the spectral windows analyzed below are treated as phenomenological observables rather than being uniquely assigned to specific microscopic mechanisms. For clarity in the discussion below, we refer to these regions as the edge-shift window and the state filling window, respectively. 

Figure~\ref{fig:SM_shortScan}(b) shows temporal responses extracted from the spectral regions indicated in Fig.~\ref{fig:SM_shortScan}(a).
These traces probe the state filling region (black markers) and the edge shift region (square markers in
Fig.~\ref{fig:SM_shortScan}(b)). Both features exhibit an instrument-response limited rise. Importantly, both responses rise much faster than the coherent lattice motion discussed below (phonon period $\sim$188~fs), indicating that the early-time dynamics are predominantly electronic in origin rather than driven by structural displacement. Furthermore, the state filling signal displays a small negative contribution at negative pump--probe delays (XUV pulse preceding the IR pump). We attribute this feature to the response of the A, B and A$'$ core-excitons, well established at the Sn N$_{4,5}$ edge in SnS \cite{Taniguchi1990}, populated by the XUV transition and subsequently dephased by the IR pulse, similarly to previous observations in WS$_2$ \cite{Chang2021}.

To quantitatively describe these dynamics, we model the rise of the differential absorption in both the state filling and edge-shift regions using exponentially modified Gaussian functions. For the state filling region, an additional term is included to account for the decay of the core-exciton contribution toward negative delays. The resulting fits are shown in Fig.~\ref{fig:SM_shortScan}(b) as the red and blue curves, respectively. Figure~\ref{fig:SM_shortScan}(c) shows the fitted decomposition of the state filling response into the core-exciton and conduction-band filling contributions, shown as gold and purple curves, respectively. From this fit, we extract a core-exciton coherence time of $13 \pm 2$ fs for the A$'$ exciton, comparable to the $\sim$10 fs coherence time previously reported for WS$_2$ \cite{Chang2021}. The B core-exciton exhibits a similar coherence time of $10 \pm 2$~fs.

\subsection{Density-dependent recombination dynamics}

\begin{figure*}[htbp!]
    \centering
    \includegraphics[width=1\linewidth]{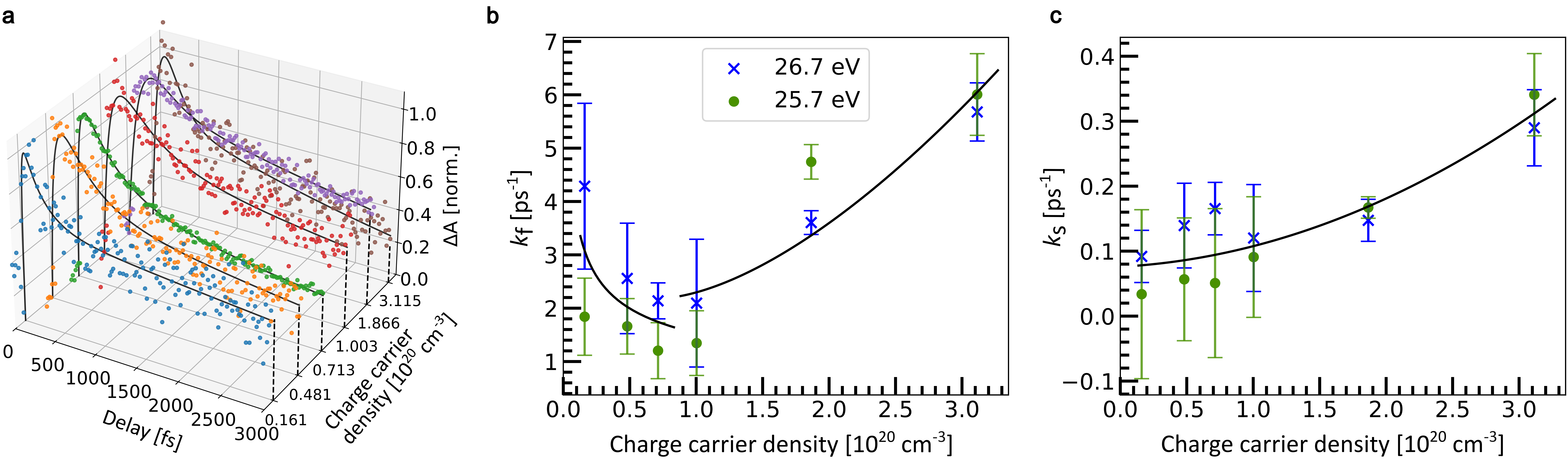}
    \caption{(a) Charge carrier density-dependent temporal responses extracted from the state filling window centered at 26.7 eV (colored points) and corresponding fits (solid black curves). $z$-axis shows normalized differential absorption, $\Delta A$. Linear offset along density axis for clarity. (b,c) Decay rate fit constants as a function of charge carrier density. Blue cross-hatches: decay rates from 26.7 eV state filling window. Green circles: decay rates from 25.7 eV sate filling window Error bars indicate 1 standard deviation. (b) shows fast decay rate $k_\text{f}$. Solid curves show power law fits to the two cooling regimes. (c) shows slow decay rate $k_\text{s}$. Solid curve shows fit of the decay rates extracted from the 25.7 and 26.7 state filling windows using Eq. \ref{eq:Gamma_def}. }
    \label{fig:fig_3_lout_fits}
\end{figure*}

We now investigate the temporal dynamics of the spin-orbit split CB spectral regions at 25.5 eV and 26.5 eV as function of carrier density. In Fig.~\ref{fig:fig_3_lout_fits}(a), we present the conduction band decay as a function of delay, for an integrated energy region around 26.5 eV (colored points), for a set of pump induced carrier densities. We fit the CB response to the following function,
\begin{equation}
    \Delta \mathrm{A_{CB}}(\tau) = \left( 1 - e ^ {-k_r \tau} \right) \left( A_f e ^ {-k_f \tau}  + A_s e ^ {-k_s \tau} \right),
\end{equation}
where the first bracket accounts for the rise of the signal with associated rate $k_r$, and the second bracket describes a bi-exponential decay consisting of a ``fast'' ($k_f$) and a ``slow'' ($k_s$) rate, with associated amplitudes $A_f$ and $A_s$, respectively. The resulting fits (black curves) are shown in Fig.~\ref{fig:fig_3_lout_fits}(a) as function of carrier density, obtained by varying the power density of the optical excitation laser pulse. The same procedure is repeated for the $4d_{5/2}\rightarrow$CB transition. Fig.~\ref{fig:fig_3_lout_fits}(b,c), shows the fast and slow decay rates ($k_f$ and $k_s$), as function of carrier density, extracted from the fits for the 4d$_{3/2}\rightarrow\text{CB}$ (blue cross-hatches) and 4d$_{5/2}\rightarrow\text{CB}$ (green points) transitions. The error bars are calculated from the fit error and are indicated as $\pm1$ standard deviation. The solid black curves show the fit of the decay rates extracted from the 25.7 and 26.7 eV state filling windows. 

In both Fig.~\ref{fig:fig_3_lout_fits} (b) and (c), we distinguish two different regimes, slowly varying rates up to a carrier density of $n =1.0\times10^{20}$~cm$^{-3}$, and increasing rates above. In keeping with previous analyses, we attribute the slow decay rates ($k_s$) with characteristic decay time $11\pm4.9$ ps to carrier recombination \cite{Li2015}, while the fast decay rates ($k_f$), with decay times on average equal to $470\pm180$ fs are associated with carrier cooling \cite{Alberding2016}.

To quantify the recombination dynamics underlying the slow decay component, we analyze the carrier-density dependence of the extracted recombination rate $k_s(n)\equiv\Gamma_\text{rec}(n)$ within a rate-equation framework. We define the effective recombination rate per carrier as \cite{Senty2015},
\begin{equation}
\Gamma_\text{rec}(n)
\equiv
-\frac{1}{n}\frac{\mathrm{d}n}{\mathrm{d}t}
=
\frac{1}{n}\left(
k_1 n
+
k_2 n^2
+
k_3 n^3
\right)
=
k_1 + k_2 n + k_3 n^2 ,
\label{eq:Gamma_def}
\end{equation}
where $k_1$, $k_2$, and $k_3$ parameterize Shockley-Read-Hall, radiative, and Auger recombination channels, respectively. Fitting Eq.~\eqref{eq:Gamma_def} to the experimentally extracted slow decay rates $\Gamma(n)=k_s(n)$ yields,
\begin{align}
k_1 &= 7.60\times10^{10}\ \mathrm{s^{-1}}, \nonumber\\
k_2 &= 1.09\times10^{-10}\ \mathrm{cm^{3}\,s^{-1}}, \nonumber\\
k_3 &= 2.08\times10^{-30}\ \mathrm{cm^{6}\,s^{-1}} .
\label{eq:Gamma_fit_values}
\end{align}
The magnitude of $k_1$ corresponds to an effective low-density recombination lifetime of $\tau_1 = k_1^{-1} \approx 13$~ps. This time scale is typical of interface-limited recombination in semiconductor nanostructures \cite{Li2015}. The extracted low-density recombination lifetime, $\tau_1$, is naturally explained if carrier recombination in the present ultrathin $\pi$-SnS films is governed by interfaces rather than by bulk processes. In a thin semiconductor where carrier transport to recombination-active interfaces is fast compared to intrinsic bulk recombination, the population decay is controlled primarily by an effective interface (and surface) recombination velocity and is therefore expected to be only weakly dependent on the bulk carrier density. The approximately density-independent slow decay observed at low excitation densities is fully consistent with this interface-limited regime, which is expected to dominate in $\sim$30~nm thick films where the surface-to-volume ratio is high. Microscopically, structural discontinuities and defect states at the free surface, buried interface, and internal grain boundaries introduce localized mid-gap states that act as efficient carrier traps, mediating electron--hole recombination through a Shockley--Read--Hall mechanism.

Across the experimentally accessible carrier-density range, the bimolecular term $k_2 n$ remains negligible compared to $k_1$, indicating that radiative recombination does not make a significant contribution to the observed dynamics. In contrast, the quadratic term $k_3 n^2$ becomes comparable to $k_1$ at carrier densities above $\sim1\times10^{20}$~cm$^{-3}$, indicating the onset of Auger-dominated recombination. The extracted Auger coefficient $k_3$ is consistent with previously reported values for Ge ($k_3 = 1 \times 10^{-30}$ cm$^6$s$^{-1}$) \cite{Li2015}. 

\subsection{Density-dependent cooling dynamics}
\label{subsec:density_cooling}

To interpret the carrier-density dependence of the experimentally extracted cooling rate ($k_f$), we start from a two-temperature model (2TM) describing the coupled dynamics of the electronic and lattice subsystems. For small perturbations about equilibrium and assuming slowly evolving $\Delta T_L$, the electronic temperature evolves as \cite{Caruso2022},
\begin{equation}
C_e(n)\,\frac{\mathrm{d}\Delta T_e}{\mathrm{d}t}
=
-\,G\,\bigl(\Delta T_e-\Delta T_L\bigr)
- P_{\mathrm{cc}}(n,\Delta T_e),
\label{eq:2TM_linear}
\end{equation}
where $C_e(n)$ is the electronic heat capacity, $G$ is the electron--phonon coupling constant, $\Delta T_L$ is the lattice temperature, and $P_{\mathrm{cc}}$ accounts for additional carrier--carrier (many-body) energy-loss pathways that become significant at high photoexcited carrier densities. Experimentally, we extract an effective cooling rate, denoted $\Gamma_{\mathrm{cool}}(n)$, from the decay of the conduction-band state filling signal in $\Delta A(\tau,\epsilon)$. We decompose this rate as $\Gamma_{\mathrm{cool}}(n)=\Gamma_{\mathrm{ph}}(n)+\Gamma_{\mathrm{Auger}}(n)$, where $\Gamma_{\mathrm{ph}}$ denotes the low-density cooling channel limited by electron--phonon coupling and electronic heat capacity, and $\Gamma_{\mathrm{Auger}}$ an additional carrier--carrier contribution that becomes relevant at high carrier densities.

\paragraph*{Low-density regime: heat-capacity--limited cooling.}
In the absence of a strong density-dependent $P_{\mathrm{cc}}$ term, the 2TM predicts an effective cooling rate,
\begin{equation}
\Gamma_{\mathrm{ph}}(n) \simeq \frac{G}{C_e(n)}.
\label{eq:k_G_over_Ce}
\end{equation}
For a degenerate or partially degenerate carrier gas with a parabolic band, the electronic carrier density is proportional to $C_e(n)\propto g(E_F)\,T_e$, and the density of states at the Fermi level scales as $g(E_F)\propto n^{1/3}$. Under the assumption of carrier density invariant electron phonon coupling, this implies,
\begin{equation}
C_e(n)\propto n^{1/3}
\quad\Rightarrow\quad
\Gamma_{\mathrm{ph}}(n)\propto n^{-1/3}.
\label{eq:Ce_scaling}
\end{equation}
We quantify this dependence by fitting the first three (low-density) data points to the power law,
\begin{equation}
\Gamma_{\mathrm{ph}}(n)=a_{\mathrm{ph}}\,n^{b_{\mathrm{ph}}},
\label{eq:k1_fit}
\end{equation}
which yields $a_{\mathrm{ph}} = 6.64\pm 2.9\times10^7$ cm$^3$ ps$^{-1}$ and $b_{\mathrm{ph}}=-0.3818\pm0.0430$. The close agreement between the measured exponent and the expected $-1/3$ scaling indicates that, in the low-density regime, the observed decrease of the effective cooling rate is governed primarily by the growth of the electronic heat capacity rather than by a reduction of the microscopic electron--phonon coupling strength or a hot-phonon bottleneck.

\paragraph*{High-density regime: Auger-assisted (carrier--carrier) cooling.}

At high photoexcited carrier densities, the experimentally extracted cooling rate deviates from the low-density trend and increases strongly with carrier density. To isolate this additional cooling contribution, we subtract the low-density component $\Gamma_{\mathrm{ph}}(n)$ from the full dataset and fit the residual (high-density) cooling rate to a power law,
\begin{equation}
\Gamma_{\mathrm{Auger}}(n)=a_{\mathrm{Auger}}\,n^{b_{\mathrm{Auger}}},
\label{eq:Gamma_Auger_fit}
\end{equation}
yielding an exponent $b_{\mathrm{Auger}} = 1.66\pm0.39$. This scaling is incompatible with phonon-mediated cooling alone and indicates the emergence of a carrier--carrier dominated energy relaxation mechanism.

In the high-density regime, the accelerated cooling suggests that an additional many-body relaxation channel becomes active. Notably, this is the same carrier-density regime where our recombination analysis indicates that Auger recombination becomes competitive with the density-independent interface term, suggesting that strong Coulomb-mediated three-carrier scattering is already important and can naturally also enable Auger-assisted intraband energy redistribution. We hypothesize that, once carrier--carrier interactions become strong, Coulomb-mediated electron--hole scattering can transfer energy from hot electrons to holes without immediate carrier annihilation (Auger cooling) \cite{Efros1995,Klimov1998,Melnychuk2021}. If holes subsequently dissipate energy efficiently via phonon emission, the energy transfer from electrons to holes followed by phonon emission of the holes would increase the net electronic energy-loss rate and thus appear as faster electron cooling in the transient absorption observable.

Within the two-temperature model, the effective cooling rate is determined by the ratio of the relevant electronic energy-loss channel to the electronic heat capacity. The rate of such energy-exchange events is assumed to scale with the number of available interacting electron--hole pairs,
\begin{equation}
P_{\mathrm{cc}}(n,\Delta T_e) \propto n_e n_h\,\Delta T_e \sim n^2\,\Delta T_e ,
\end{equation}
where $P_{\mathrm{cc}}$ denotes the effective carrier--carrier cooling power and $\Delta T_e$ is the change in electronic temperature, and the number of excited electrons, $n_e$, is assumed to be similar to the number of excited holes, $n_h$. As discussed above, for a degenerate or partially degenerate carrier gas, the electronic heat capacity scales as $C_e(n)\propto n^{1/3}$. The resulting Auger-assisted cooling
rate therefore follows,
\begin{equation}
\Gamma_{\mathrm{Auger}}(n)
\sim
\frac{P_{\mathrm{cc}}(n,\Delta T_e)}{C_e(n)\,\Delta T_e}
\propto
\frac{n^2}{n^{1/3}}
=
n^{5/3}.
\label{eq:auger_cool_scaling}
\end{equation}
The experimentally extracted exponent $b_{\mathrm{Auger}} = 1.66\pm0.39$ is in excellent agreement with the expected $n^{5/3}$ scaling, supporting that Auger-assisted carrier--carrier energy exchange governs the enhanced cooling dynamics at high photoexcited carrier densities. Taken together, these results reveal a crossover from a low-density regime governed by phonon-limited, heat capacity dominated cooling to a high-density regime in which Auger-assisted carrier--carrier scattering provides an additional, strongly density-dependent cooling pathway, establishing a direct connection between the phenomenological cooling kinetics extracted from the experiment and the microscopic many-body interactions that govern high-density carrier dynamics in $\pi$-SnS.

\subsection{Coherent phonon motion}

Interband optical excitation by the NIR pump impulsively perturbs the electronic polarization and drives a coherent Raman force, launching lattice wavepackets via impulsive stimulated Raman scattering (ISRS). In the attosecond transient absorption measurements, this lattice response manifests as coherent phonon oscillations superimposed on the carrier-induced BGR signal.
\begin{figure}[htbp!]
    \centering
    \includegraphics[width=0.75\linewidth]{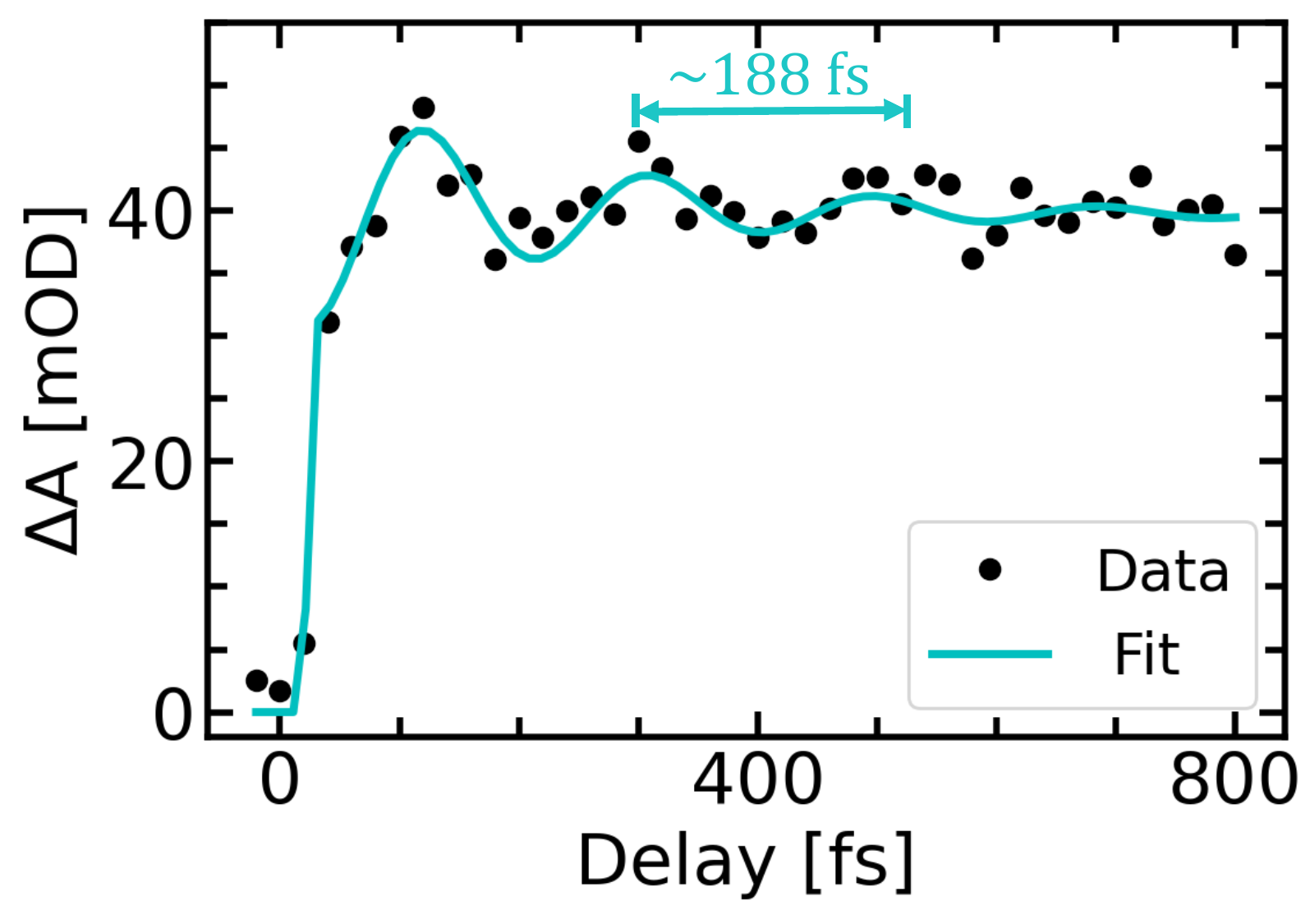}
    \caption{Coherent phonon oscillations as function of delay averaged over the spectral region between 26.25 and 26.35 eV with strong lattice response. The data are shown in black while the magenta curve is the fit to the data using Eq.~\ref{eq:eq_ph}.}
    \label{fig:fig_4}
\end{figure}
At a photoexcited carrier density of $7.1\times10^{19}$~cm$^{-3}$, weak oscillations are observed in the edge-shift region centered at 26.3~eV, as shown in Fig.~\ref{fig:fig_2}. This spectral region is particularly sensitive to lattice motion due to changes of the conduction band energy, through electron-phonon coupling, or via changes to the level core-energy. Fig.~\ref{fig:fig_4} shows an energy-integrated lineout around this feature, where coherent phonon oscillations are clearly resolved. We extract the phonon frequency $\Omega$, phase $\varphi$, and dephasing rate $\gamma$ by fitting the transient signal to,
\begin{equation}
\begin{split}
\Delta \mathrm{A_{phonon}}(\tau)& =f(\tau-t_0)\\& * [Ae^{-\beta \tau}\!\! - Be^{-\gamma \tau}\!\! \cos{( \Omega \tau - \varphi)}]
\label{eq:eq_ph}
\end{split}
\end{equation}
where $f(\tau-t_0)$ is the instrument response function, $A$ is an amplitude factor, $\beta$ accounts for the initial buildup of the state filling signal and $\gamma$ describes phonon dephasing. The resulting fit is shown in Fig.~\ref{fig:fig_4}.

From the fit, we obtain a phonon period of $188\pm6.8$~fs, corresponding to a frequency of $177$~cm$^{-1}$. Raman spectroscopy measurements of $\pi$-SnS, shown in Fig.~\ref{fig:SI_SnSpi} in the SM, reveal a dense spectrum of phonon modes, consistent with the large primitive cell of $\pi$-SnS containing 64 atoms and thus 192 vibrational modes \cite{Skelton2017_2,Guc20201,Abutbul2016}. 
The most intense Raman modes are observed in a spectral range around our retrieved value (160-190~cm$^{-1}$). Density-functional-theory calculations predict several $\pi$-SnS optical modes within this range \cite{Skelton2017_2,Guc20201}, including modes in close agreement with the extracted frequency, precluding a unique mode assignment based on frequency alone.

The extracted phonon dephasing time is $229$~fs, indicating a rapid loss of phase coherence. Such fast dephasing is expected in $\pi$-SnS due to the dense manifold of available optical phonon modes, which could enable efficient anharmonic phonon--phonon scattering and provides multiple dephasing channels. The fitted phase, $\varphi = 0.69\pi \pm 0.07\pi$, corresponds to a sine-like oscillation launched at time zero, consistent with the impulsive limit of impulsive stimulated Raman scattering (ISRS) \cite{Zeiger1992}. This behavior indicates that the coherent lattice motion is driven by an impulsive Raman force rather than by a displacive shift of the equilibrium lattice configuration.

\section{Summary}

In summary, we use attosecond transient absorption spectroscopy to resolve carrier and lattice dynamics in metastable cubic $\pi$-SnS with element- and orbital-specific sensitivity. Following femtosecond near-infrared excitation, the Sn $4d\rightarrow$CBM response exhibits a characteristic transient lineshape consisting of conduction-band filling together with pump-induced edge shifts driven by carrier-induced renormalization (including band-gap and core-exciton contributions). By analyzing well-defined spectral windows of the response over a wide range of photoexcited carrier densities, we extract two characteristic timescales: a fast component associated with intraband carrier cooling and a slow component associated with interband recombination.

At low excitation densities, the slow decay is approximately independent of carrier density and is consistent with interface-limited recombination, which is expected to dominate in the thin ($\sim$30~nm thick) films studied here. With increasing carrier density, both the recombination and cooling rates accelerate, marking a crossover to a multicarrier kinetic regime. The correlated density dependence of the slow recombination component and the fast cooling component indicates that carrier--carrier interactions become operative above a threshold density, consistent with the onset of Auger-dominated pathways and Auger-assisted intraband energy redistribution that appears as an enhanced cooling rate in the transient absorption observable.

In addition to carrier dynamics, the XUV response reveals coherent phonon oscillations with a period of $188\pm6.8$~fs, demonstrating coupling between electronic excitation and lattice motion in $\pi$-SnS. Together, these results provide a unified picture of how surface effects, multicarrier interactions, and coherent lattice dynamics jointly govern nonequilibrium behavior in this emerging semiconductor. More broadly, our work highlights the capability of attosecond transient absorption spectroscopy to disentangle competing ultrafast processes in structurally complex solids and establishes $\pi$-SnS as a promising platform for exploring high-density carrier dynamics relevant to optoelectronic and energy-conversion applications.

\section{Acknowledgements}
HL would like to acknowledge fruitful discussion with Hendrik Utzat. This work was supported by the Department of Energy, Office of Science, Basic Energy Science (BES) Program within the Materials Science and Engineering Division (contract DE-AC02-05CH11231), through the Fundamentals of Semiconductor Nanowires Program through the Lawrence Berkeley National Laboratory. Support for laser instrumentation, vacuum hardware, and additional personel is from AFOSR grant numbers FA9550-19-1-0314, FA9550-24-1-0184 (K.J.G.), FA9550-20-0334 (R.Q.B.) and FA9550-22-1-0451 (DURIP). H.L. acknowledges support from the Swedish Research Council (2023-06502) and the Sweden-America Foundation. J.R.A acknowledges the NSF GRFP under grant No DGE 2146752. H.K.D.L acknowledges support from the National Science Foundation’s Graduate Research Fellowship Program (NSF GRFP) under grant DGE 1752814. Work at the Molecular Foundry was supported by the Office of Science, Office of Basic Energy Sciences, of the U.S. Department of Energy under Contract No. DE-AC02-05CH11231. N.O. and T.K.J. acknowledge support for travel and accommodation from the Royal Physiographic Society of Lund (YR-PhD:45993) and the NanoLund New Technique Fund. N.O. acknowledges support from the European Research Council (Grants No. 884900, No 851201).

\section{Author contributions}
H. L., K. X., N. O., T. K. J., contributed equally to this work. H. L., S. R. L. and C. H. conceptualized the work. K. X. led the experiment with the support of N. O., T. K. J., H. L., R. Q.-B. and J. R. A.. N. O., T. K. J., and H. L. analyzed the data. All authors contributed to the scientific discussion. H. L., K. X., N. O., and T. K. J., wrote the manuscript with feedback and comments from all authors. C. H. synthesized the samples. C. H., L. V. and H. K. L. D. performed characterization of the samples. 

\bibliography{Ref_lib}

@article{Ahmet2019,
	author = {Ahmet, Ibbi Y. and Guc, Maxim and S{\'a}nchez, Yudania and Neuschitzer, Markus and Izquierdo-Roca, Victor and Saucedo, Edgardo and Johnson, Andrew L.},
	doi = {10.1039/C9RA01938C},
	issue = {26},
	journal = {RSC Adv.},
	pages = {14899-14909},
	publisher = {The Royal Society of Chemistry},
	title = {Evaluation of AA-CVD deposited phase pure polymorphs of SnS for thin films solar cells},
	url = {http://dx.doi.org/10.1039/C9RA01938C},
	volume = {9},
	year = {2019}}

@article{Polivtseva2017,
	author = {Polivtseva, S. and Katerski, A. and K{\"a}rber, E. and Oja Acik, I. and Mere, A. and Mikli, V. and Krunks, M.},
	date = {2017/07/01/},
	doi = {https://doi.org/10.1016/j.tsf.2017.01.014},
	isbn = {0040-6090},
	journal = {Thin Solid Films},
	journal1 = {E-MRS 2016 Spring Meeting, Symposium V, Thin-Film Chalcogenide Photovoltaic Materials},
	pages = {179--184},
	title = {Post-deposition thermal treatment of sprayed SnS films},
	url = {https://www.sciencedirect.com/science/article/pii/S0040609017300147},
	volume = {633},
	year = {2017}}

@article{Abutbul2020,
	author = {Abutbul, Ran E. and Golan, Yuval},
	doi = {10.1039/D0CE00797H},
	issue = {37},
	journal = {CrystEngComm},
	pages = {6170-6181},
	publisher = {The Royal Society of Chemistry},
	title = {Chemical epitaxy of π-phase cubic tin monosulphide},
	url = {http://dx.doi.org/10.1039/D0CE00797H},
	volume = {22},
	year = {2020}}

@article{Efros1995,
	author = {Al. L. Efros and V.A. Kharchenko and M. Rosen},
	doi = {https://doi.org/10.1016/0038-1098(94)00760-8},
	issn = {0038-1098},
	journal = {Solid State Communications},
	number = {4},
	pages = {281-284},
	title = {Breaking the phonon bottleneck in nanometer quantum dots: Role of Auger-like processes},
	url = {https://www.sciencedirect.com/science/article/pii/0038109894007608},
	volume = {93},
	year = {1995}}

@article{Klimov1998,
	author = {Klimov, Victor I. and McBranch, Duncan W.},
	doi = {10.1103/PhysRevLett.80.4028},
	issue = {18},
	journal = {Phys. Rev. Lett.},
	month = {May},
	numpages = {0},
	pages = {4028--4031},
	publisher = {American Physical Society},
	title = {Femtosecond $1\mathit{P}$-to- $1\mathit{S}$ Electron Relaxation in Strongly Confined Semiconductor Nanocrystals},
	url = {https://link.aps.org/doi/10.1103/PhysRevLett.80.4028},
	volume = {80},
	year = {1998}}

@article{Rossi2025,
	author = {Rossi, Thomas C. and Qiao, Lu and Dykstra, Conner P. and Rodrigues Pela, Ronaldo and Gnewkow, Richard and Wallick, Rachel F. and Burke, John H. and Nicholas, Erin and March, Anne Marie and Doumy, Gilles and Buchholz, D. Bruce and Deparis, Christiane and Z{\'u}{\~n}iga-P{\'e}rez, Jes{\'u}s and Weise, Michael and Ellmer, Klaus and Fondell, Mattis and Draxl, Claudia and van der Veen, Renske M.},
	date = {2025/08/20},
	doi = {10.1038/s43246-025-00909-w},
	id = {Rossi2025},
	isbn = {2662-4443},
	journal = {Communications Materials},
	number = {1},
	pages = {191},
	title = {Dynamic control of X-ray core-exciton resonances by Coulomb screening in photoexcited semiconductors},
	url = {https://doi.org/10.1038/s43246-025-00909-w},
	volume = {6},
	year = {2025}}

@article{Bilousov2017,
	author = {Bilousov, Oleksandr V. and Ren, Yi and T{\"o}rndahl, Tobias and Donzel-Gargand, Olivier and Ericson, Tove and Platzer-Bj{\"o}rkman, Charlotte and Edoff, Marika and H{\"a}gglund, Carl},
	date = {2017/04/11},
	doi = {10.1021/acs.chemmater.6b05323},
	isbn = {0897-4756},
	journal = {Chemistry of Materials},
	journal1 = {Chemistry of Materials},
	journal2 = {Chem. Mater.},
	month = {04},
	number = {7},
	pages = {2969--2978},
	publisher = {American Chemical Society},
	title = {Atomic Layer Deposition of Cubic and Orthorhombic Phase Tin Monosulfide},
	type = {doi: 10.1021/acs.chemmater.6b05323},
	url = {https://doi.org/10.1021/acs.chemmater.6b05323},
	volume = {29},
	year = {2017},
	year1 = {2017}}

@article{Taniguchi1990,
	author = {Taniguchi, M. and Johnson, R. L. and Ghijsen, J. and Cardona, M.},
	doi = {10.1103/PhysRevB.42.3634},
	issue = {6},
	journal = {Phys. Rev. B},
	month = {Aug},
	numpages = {0},
	pages = {3634--3643},
	publisher = {American Physical Society},
	title = {Core excitons and conduction-band structures in orthorhombic GeS, GeSe, SnS, and SnSe single crystals},
	url = {https://link.aps.org/doi/10.1103/PhysRevB.42.3634},
	volume = {42},
	year = {1990}}

@article{Chang2021,
	author = {Chang, Hung-Tzu and Guggenmos, Alexander and Chen, Christopher T. and Oh, Juwon and G\'eneaux, Romain and Chuang, Yi-De and Schwartzberg, Adam M. and Aloni, Shaul and Neumark, Daniel M. and Leone, Stephen R.},
	doi = {10.1103/PhysRevB.104.064309},
	issue = {6},
	journal = {Phys. Rev. B},
	month = {Aug},
	numpages = {14},
	pages = {064309},
	publisher = {American Physical Society},
	title = {Coupled valence carrier and core-exciton dynamics in ${\mathrm{WS}}_{2}$ probed by few-femtosecond extreme ultraviolet transient absorption spectroscopy},
	url = {https://link.aps.org/doi/10.1103/PhysRevB.104.064309},
	volume = {104},
	year = {2021}}

@article{Banai2014,
	author = {Banai, R. E. and Burton, L. A. and Choi, S. G. and Hofherr, F. and Sorgenfrei, T. and Walsh, A. and To, B. and Cr{\"o}ll, A. and Brownson, J. R. S.},
	doi = {10.1063/1.4886915},
	eprint = {},
	issn = {0021-8979},
	journal = {Journal of Applied Physics},
	month = {07},
	number = {1},
	pages = {013511},
	title = {Ellipsometric characterization and density-functional theory analysis of anisotropic optical properties of single-crystal α-SnS},
	url = {https://doi.org/10.1063/1.4886915},
	volume = {116},
	year = {2014}}

@article{Sine2011,
	author = {Sinsermsuksakul, Prasert and Heo, Jaeyeong and Noh, Wontae and Hock, Adam S. and Gordon, Roy G.},
	doi = {https://doi.org/10.1002/aenm.201100330},
	eprint = {},
	journal = {Advanced Energy Materials},
	number = {6},
	pages = {1116-1125},
	title = {Atomic Layer Deposition of Tin Monosulfide Thin Films},
	url = {https://advanced.onlinelibrary.wiley.com/doi/pdf/10.1002/aenm.201100330},
	volume = {1},
	year = {2011}}

@article{Jaramillo2016,
	author = {Jaramillo, R. and Sher, Meng-Ju and Ofori-Okai, Benjamin K. and Steinmann, V. and Yang, Chuanxi and Hartman, Katy and Nelson, Keith A. and Lindenberg, Aaron M. and Gordon, Roy G. and Buonassisi, T.},
	doi = {10.1063/1.4940157},
	eprint = {},
	issn = {0021-8979},
	journal = {Journal of Applied Physics},
	month = {01},
	number = {3},
	pages = {035101},
	title = {Transient terahertz photoconductivity measurements of minority-carrier lifetime in tin sulfide thin films: Advanced metrology for an early stage photovoltaic material},
	url = {https://doi.org/10.1063/1.4940157},
	volume = {119},
	year = {2016}}

@article{Schultze2014,
	author = {Schultze, Martin and Ramasesha, Krupa and Pemmaraju, C. D. and Sato, S. A. and Whitmore, D. and Gandman, A. and Prell, James S. and Borja, L. J. and Prendergast, D. and Yabana, K. and Neumark, Daniel M. and Leone, Stephen R.},
	date = {2014/12/12},
	doi = {10.1126/science.1260311},
	journal = {Science},
	journal1 = {Science},
	journal2 = {Science},
	month = {2026/02/05},
	n2 = {Electron transfer from valence to conduction band states in semiconductors is the basis of modern electronics. Here, attosecond extreme ultraviolet (XUV) spectroscopy is used to resolve this process in silicon in real time. Electrons injected into the conduction band by few-cycle laser pulses alter the silicon XUV absorption spectrum in sharp steps synchronized with the laser electric field oscillations. The observed \~{}450-attosecond step rise time provides an upper limit for the carrier-induced band-gap reduction and the electron-electron scattering time in the conduction band. This electronic response is separated from the subsequent band-gap modifications due to lattice motion, which occurs on a time scale of 60 $\pm$10 femtoseconds, characteristic of the fastest optical phonon. Quantum dynamical simulations interpret the carrier injection step as light-field?induced electron tunneling. Excited electrons in semiconducting silicon are tracked on a time scale faster than the lattice vibrations. {$[$}Also see Perspective by Spielmann{$]$} The ultimate speed limit in electronic circuitry is set by the motion of the electrons themselves. Schultze et al. applied attosecond spectroscopy to glimpse this motion in a sample of silicon, the semiconducting building block of modern integrated circuits (see the Perspective by Spielmann). The technique distinguished the electron dynamics?which proceed faster than a quadrillionth of a second after laser excitation?from the comparatively slower lattice motion of the silicon atomic nuclei. Science, this issue p. 1348; see also p. 1293},
	number = {6215},
	pages = {1348--1352},
	publisher = {American Association for the Advancement of Science},
	title = {Attosecond band-gap dynamics in silicon},
	type = {doi: 10.1126/science.1260311},
	url = {https://doi.org/10.1126/science.1260311},
	volume = {346},
	year = {2014},
	year1 = {2014}}

@article{Senty2015,
	author = {Senty, Tess R. and Cushing, Scott K. and Wang, Congjun and Matranga, Christopher and Bristow, Alan D.},
	date = {2015/03/19},
	doi = {10.1021/jp512500g},
	isbn = {1932-7447},
	journal = {The Journal of Physical Chemistry C},
	journal1 = {The Journal of Physical Chemistry C},
	journal2 = {J. Phys. Chem. C},
	month = {03},
	number = {11},
	pages = {6337--6343},
	publisher = {American Chemical Society},
	title = {Inverting Transient Absorption Data to Determine Transfer Rates in Quantum Dot--TiO2 Heterostructures},
	type = {doi: 10.1021/jp512500g},
	url = {https://doi.org/10.1021/jp512500g},
	volume = {119},
	year = {2015},
	year1 = {2015}}

@article{Caruso2022,
	author = {Caruso, Fabio and Novko, Dino},
	date = {2022/12/31},
	doi = {10.1080/23746149.2022.2095925},
	isbn = {null},
	journal = {Advances in Physics: X},
	journal1 = {Advances in Physics: X},
	journal2 = {Advances in Physics: X},
	month = {12},
	number = {1},
	pages = {2095925},
	publisher = {Taylor \& Francis},
	title = {Ultrafast dynamics of electrons and phonons: from the two-temperature model to the time-dependent Boltzmann equation},
	type = {doi: 10.1080/23746149.2022.2095925},
	url = {https://doi.org/10.1080/23746149.2022.2095925},
	volume = {7},
	year = {2022},
	year1 = {2022}}

@article{Padova1994,
	author = {De Padova, P. and Fanfoni, M. and Larciprete, R. and Mangiantini, M. and Priori, S. and Perfetti, P.},
	date = {1994/07/01/},
	doi = {https://doi.org/10.1016/0039-6028(94)90058-2},
	isbn = {0039-6028},
	journal = {Surface Science},
	number = {3},
	pages = {379--391},
	title = {A synchrotron radiation photoemission study of the oxidation of tin},
	url = {https://www.sciencedirect.com/science/article/pii/0039602894900582},
	volume = {313},
	year = {1994}}

@article{Kishore2023,
	author = {Kishore Bhat, T. R. and Jeganath, K. and George, Sajan D. and Raviprakash, Y.},
	date = {2023/03/13},
	doi = {10.1007/s10854-023-10157-8},
	id = {Kishore Bhat2023},
	isbn = {1573-482X},
	journal = {Journal of Materials Science: Materials in Electronics},
	number = {8},
	pages = {747},
	title = {Annealing-induced phase conversion on spray pyrolyzed cubic-SnS thin films},
	url = {https://doi.org/10.1007/s10854-023-10157-8},
	volume = {34},
	year = {2023}}

@article{Hegde2020,
	author = {Hegde, S. S. and Murahari, Prashantha and Fernandes, Brian Jeevan and Venkatesh, R. and Ramesh, K.},
	date = {2020/04/15/},
	doi = {https://doi.org/10.1016/j.jallcom.2019.153116},
	isbn = {0925-8388},
	journal = {Journal of Alloys and Compounds},
	pages = {153116},
	title = {Synthesis, thermal stability and structural transition of cubic SnS nanoparticles},
	url = {https://www.sciencedirect.com/science/article/pii/S0925838819343622},
	volume = {820},
	year = {2020}}

@article{Abutbul2016_2,
	author = {Abutbul, R. E. and Garcia-Angelmo, A. R. and Burshtein, Z. and Nair, M. T. S. and Nair, P. K. and Golan, Y.},
	doi = {10.1039/C6CE00647G},
	issue = {27},
	journal = {CrystEngComm},
	pages = {5188-5194},
	publisher = {The Royal Society of Chemistry},
	title = {Crystal structure of a large cubic tin monosulfide polymorph: an unraveled puzzle},
	url = {http://dx.doi.org/10.1039/C6CE00647G},
	volume = {18},
	year = {2016}}

@article{Bilousov2021,
	author = {Bilousov, Oleksandr V. and Voznyi, Andrii and Landeke-Wilsmark, Bj{\"o}rn and Villamayor, Michelle Marie S. and Nyberg, Tomas and H{\"a}gglund, Carl},
	date = {2021/04/27},
	doi = {10.1021/acs.chemmater.1c00241},
	isbn = {0897-4756},
	journal = {Chemistry of Materials},
	journal1 = {Chemistry of Materials},
	journal2 = {Chem. Mater.},
	month = {04},
	number = {8},
	pages = {2901--2912},
	publisher = {American Chemical Society},
	title = {Substrate Effects on Crystal Phase in Atomic Layer Deposition of Tin Monosulfide},
	type = {doi: 10.1021/acs.chemmater.1c00241},
	url = {https://doi.org/10.1021/acs.chemmater.1c00241},
	volume = {33},
	year = {2021},
	year1 = {2021}}

@article{Peng2025,
	author = {Peng, Zanxiong and Cong, Borong and Cao, Jiajun and Li, Chunlian and Shen, Xiaodong and Liang, Weizheng},
	doi = {10.1039/D5CP00821B},
	issue = {23},
	journal = {Phys. Chem. Chem. Phys.},
	pages = {12267-12273},
	publisher = {The Royal Society of Chemistry},
	title = {Anisotropic ultrafast hot carrier dynamics of two-dimensional SnS single crystals},
	url = {http://dx.doi.org/10.1039/D5CP00821B},
	volume = {27},
	year = {2025}}

@article{Fu2017,
	author = {Fu, Jianhui and Xu, Qiang and Han, Guifang and Wu, Bo and Huan, Cheng Hon Alfred and Leek, Meng Lee and Sum, Tze Chien},
	date = {2017/11/03},
	doi = {10.1038/s41467-017-01360-3},
	id = {Fu2017},
	isbn = {2041-1723},
	journal = {Nature Communications},
	number = {1},
	pages = {1300},
	title = {Hot carrier cooling mechanisms in halide perovskites},
	url = {https://doi.org/10.1038/s41467-017-01360-3},
	volume = {8},
	year = {2017}}

@article{Wertheim1989,
	author = {Wertheim, G. K. and Buchanan, D. N. E.},
	date = {1989/02/01/},
	doi = {https://doi.org/10.1016/0038-1098(89)90063-X},
	isbn = {0038-1098},
	journal = {Solid State Communications},
	number = {6},
	pages = {689--692},
	title = {Crystal field splitting of core levels in β-Sn},
	url = {https://www.sciencedirect.com/science/article/pii/003810988990063X},
	volume = {69},
	year = {1989}}

@article{Drescher2025,
	author = {Drescher, Lauren B. and de Roulet, Bethany R. and Phang, Yoong Sheng and Leone, Stephen R.},
	date = {2025/09/09/},
	day = {09},
	doi = {10.1103/m5m2-7z57},
	id = {10.1103/m5m2-7z57},
	j1 = {PRB},
	journal = {Physical Review B},
	journal1 = {Phys. Rev. B},
	month = {09},
	number = {10},
	pages = {104310--},
	publisher = {American Physical Society},
	title = {Onset of coherent phonon motion in Peierls-distorted antimony by attosecond transient absorption},
	url = {https://link.aps.org/doi/10.1103/m5m2-7z57},
	volume = {112},
	year = {2025}}

@misc{Laurell2025,
	archiveprefix = {arXiv},
	author = {Hugo Laurell and Jonah R. Adelman and Elizaveta Yakovleva and Carl H{\"a}gglund and Kostiantyn Sopiha and Axel Stenquist and Han K. D. Le and Peidong Yang and Marika Edoff and Stephen R. Leone},
	eprint = {2506.05621},
	primaryclass = {cond-mat.mtrl-sci},
	title = {Coherent phonon motions and ordered vacancy compound mediated quantum path interference in Cu-poor CuIn$_{x}$Ga$_{(1-x)}$Se$_2$ (CIGS) with attosecond transient absorption},
	url = {https://arxiv.org/abs/2506.05621},
	year = {2025}}

@article{Alberding2016,
	author = {Alberding, Brian G. and Biacchi, Adam J. and Hight Walker, Angela R. and Heilweil, Edwin J.},
	date = {2016/07/21},
	doi = {10.1021/acs.jpcc.6b01684},
	isbn = {1932-7447},
	journal = {The Journal of Physical Chemistry C},
	journal1 = {The Journal of Physical Chemistry C},
	journal2 = {J. Phys. Chem. C},
	month = {07},
	number = {28},
	pages = {15395--15406},
	publisher = {American Chemical Society},
	title = {Charge Carrier Dynamics and Mobility Determined by Time-Resolved Terahertz Spectroscopy on Films of Nano-to-Micrometer-Sized Colloidal Tin(II) Monosulfide},
	type = {doi: 10.1021/acs.jpcc.6b01684},
	url = {https://doi.org/10.1021/acs.jpcc.6b01684},
	volume = {120},
	year = {2016},
	year1 = {2016}}

@article{Yang2016,
	author = {Yang, Ye and Ostrowski, David P. and France, Ryan M. and Zhu, Kai and van de Lagemaat, Jao and Luther, Joseph M. and Beard, Matthew C.},
	date = {2016/01/01},
	doi = {10.1038/nphoton.2015.213},
	id = {Yang2016},
	isbn = {1749-4893},
	journal = {Nature Photonics},
	number = {1},
	pages = {53--59},
	title = {Observation of a hot-phonon bottleneck in lead-iodide perovskites},
	url = {https://doi.org/10.1038/nphoton.2015.213},
	volume = {10},
	year = {2016}}

@article{Skelton2017_2,
	author = {Skelton, Jonathan M. and Burton, Lee A. and Oba, Fumiyasu and Walsh, Aron},
	doi = {10.1063/1.4977868},
	eprint = {},
	issn = {2166-532X},
	journal = {APL Materials},
	month = {03},
	number = {3},
	pages = {036101},
	title = {Metastable cubic tin sulfide: A novel phonon-stable chiral semiconductor},
	url = {https://doi.org/10.1063/1.4977868},
	volume = {5},
	year = {2017}}

@article{Melnychuk2021,
	author = {Melnychuk, Christopher and Guyot-Sionnest, Philippe},
	date = {2021/02/24},
	doi = {10.1021/acs.chemrev.0c00931},
	isbn = {0009-2665},
	journal = {Chemical Reviews},
	journal1 = {Chemical Reviews},
	journal2 = {Chem. Rev.},
	month = {02},
	number = {4},
	pages = {2325--2372},
	publisher = {American Chemical Society},
	title = {Multicarrier Dynamics in Quantum Dots},
	type = {doi: 10.1021/acs.chemrev.0c00931},
	url = {https://doi.org/10.1021/acs.chemrev.0c00931},
	volume = {121},
	year = {2021},
	year1 = {2021}}

@article{Abutbul2016,
	author = {Abutbul, R.E. and Segev, E. and Zeiri, L. and Ezersky, V. and Makov, G. and Golan, Y.},
	doi = {10.1039/c5ra23092f},
	journal = {RSC Advances},
	note = {Cited by: 146},
	number = {7},
	pages = {5848 -- 5855},
	publication_stage = {Final},
	source = {Scopus},
	title = {Synthesis and properties of nanocrystalline π-SnS-a new cubic phase of tin sulphide},
	type = {Article},
	url = {https://www.scopus.com/inward/record.uri?eid=2-s2.0-84955600140&doi=10.1039%2fc5ra23092f&partnerID=40&md5=43210960093fd5aa4aac1990c85e2004},
	volume = {6},
	year = {2016}}

@article{Skelton2017,
	author = {Skelton, Jonathan M. and Burton, Lee A. and Jackson, Adam J. and Oba, Fumiyasu and Parker, Stephen C. and Walsh, Aron},
	doi = {10.1039/C7CP01680H},
	issue = {19},
	journal = {Phys. Chem. Chem. Phys.},
	pages = {12452-12465},
	publisher = {The Royal Society of Chemistry},
	title = {Lattice dynamics of the tin sulphides SnS2{,} SnS and Sn2S3: vibrational spectra and thermal transport},
	url = {http://dx.doi.org/10.1039/C7CP01680H},
	volume = {19},
	year = {2017}}

@article{Li2015,
	author = {Li, Yanying and Clady, Raphael and Marshall, Ann F. and Park, Junghyun and Thombare, Shruti V. and Chan, Gerentt and Schmidt, Timothy W. and Brongersma, Mark L. and McIntyre, Paul C.},
	doi = {10.1021/acsphotonics.5b00147},
	eprint = {https://doi.org/10.1021/acsphotonics.5b00147},
	journal = {ACS Photonics},
	number = {8},
	pages = {1091-1098},
	title = {Ultrafast Carrier Dynamics of a Photo-Excited Germanium Nanowire--Air Metamaterial},
	url = {https://doi.org/10.1021/acsphotonics.5b00147},
	volume = {2},
	year = {2015}}

@article{Guc20201,
	author = {Maxim Guc and Jacob Andrade-Arvizu and Ibbi Y. Ahmet and Florian Oliva and Marcel Placidi and Xavier Alcob{\'e} and Edgardo Saucedo and Alejandro P{\'e}rez-Rodr{\'\i}guez and Andrew L. Johnson and Victor Izquierdo-Roca},
	doi = {https://doi.org/10.1016/j.actamat.2019.11.016},
	issn = {1359-6454},
	journal = {Acta Materialia},
	pages = {1-10},
	title = {Structural and vibrational properties of α- and π-SnS polymorphs for photovoltaic applications},
	url = {https://www.sciencedirect.com/science/article/pii/S135964541930744X},
	volume = {183},
	year = {2020}}

@article{Adelman2025,
	author = {Adelman, Jonah R. and Laurell, Hugo and Drescher, Lauren B. and Le, Han K. D. and Yang, Peidong and Leone, Stephen R.},
	doi = {10.1103/PhysRevB.111.184315},
	issue = {18},
	journal = {Phys. Rev. B},
	month = {May},
	numpages = {15},
	pages = {184315},
	publisher = {American Physical Society},
	title = {Coherently coupled carrier and phonon dynamics in elemental tellurium probed by extreme ultraviolet transient absorption},
	url = {https://link.aps.org/doi/10.1103/PhysRevB.111.184315},
	volume = {111},
	year = {2025}}

@article{Cushing2018,
	author = {Cushing, Scott K. and Z{\"u}rch, Michael and Kraus, Peter M. and Carneiro, Lucas M. and Lee, Angela and Chang, Hung-Tzu and Kaplan, Christopher J. and Leone, Stephen R.},
	doi = {10.1063/1.5038015},
	eprint = {},
	issn = {2329-7778},
	journal = {Structural Dynamics},
	month = {09},
	number = {5},
	pages = {054302},
	title = {{Hot phonon and carrier relaxation in Si(100) determined by transient extreme ultraviolet spectroscopy}},
	url = {https://doi.org/10.1063/1.5038015},
	volume = {5},
	year = {2018}}

@article{Klemens1966,
	author = {Klemens, P. G.},
	doi = {10.1103/PhysRev.148.845},
	issue = {2},
	journal = {Phys. Rev.},
	month = {Aug},
	numpages = {0},
	pages = {845--848},
	publisher = {American Physical Society},
	title = {Anharmonic Decay of Optical Phonons},
	url = {https://link.aps.org/doi/10.1103/PhysRev.148.845},
	volume = {148},
	year = {1966}}

@article{Attar2020,
	author = {Attar, Andrew R. and Chang, Hung-Tzu and Britz, Alexander and Zhang, Xiang and Lin, Ming-Fu and Krishnamoorthy, Aravind and Linker, Thomas and Fritz, David and Neumark, Daniel M. and Kalia, Rajiv K. and Nakano, Aiichiro and Ajayan, Pulickel and Vashishta, Priya and Bergmann, Uwe and Leone, Stephen R.},
	date = {2020/11/24},
	doi = {10.1021/acsnano.0c06988},
	isbn = {1936-0851},
	journal = {ACS Nano},
	journal1 = {ACS Nano},
	journal2 = {ACS Nano},
	month = {11},
	number = {11},
	pages = {15829--15840},
	publisher = {American Chemical Society},
	title = {Simultaneous Observation of Carrier-Specific Redistribution and Coherent Lattice Dynamics in 2H-MoTe2 with Femtosecond Core-Level Spectroscopy},
	type = {doi: 10.1021/acsnano.0c06988},
	url = {https://doi.org/10.1021/acsnano.0c06988},
	volume = {14},
	year = {2020},
	year1 = {2020}}

@article{Zeiger1992,
	author = {Zeiger, H. J. and Vidal, J. and Cheng, T. K. and Ippen, E. P. and Dresselhaus, G. and Dresselhaus, M. S.},
	doi = {10.1103/PhysRevB.45.768},
	issue = {2},
	journal = {Phys. Rev. B},
	month = {Jan},
	numpages = {0},
	pages = {768--778},
	publisher = {American Physical Society},
	title = {Theory for displacive excitation of coherent phonons},
	url = {https://link.aps.org/doi/10.1103/PhysRevB.45.768},
	volume = {45},
	year = {1992}}

@article{Palo2024,
	author = {Di Palo, N. and Inzani, G. and Dolso, G. L. and Talarico, M. and Bonetti, S. and Lucchini, M.},
	doi = {10.1063/5.0176656},
	eprint = {},
	issn = {2378-0967},
	journal = {APL Photonics},
	month = {02},
	number = {2},
	pages = {020901},
	title = {{Attosecond absorption and reflection spectroscopy of solids}},
	url = {https://doi.org/10.1063/5.0176656},
	volume = {9},
	year = {2024}}

@article{Zuerch2017,
	author = {Z{\"u}rch, Michael and Chang, Hung-Tzu and Borja, Lauren J. and Kraus, Peter M. and Cushing, Scott K. and Gandman, Andrey and Kaplan, Christopher J. and Oh, Myoung Hwan and Prell, James S. and Prendergast, David and Pemmaraju, Chaitanya D. and Neumark, Daniel M. and Leone, Stephen R.},
	date = {2017/06/01},
	doi = {10.1038/ncomms15734},
	id = {Z{\"u}rch2017},
	isbn = {2041-1723},
	journal = {Nature Communications},
	number = {1},
	pages = {15734},
	title = {Direct and simultaneous observation of ultrafast electron and hole dynamics in germanium},
	url = {https://doi.org/10.1038/ncomms15734},
	volume = {8},
	year = {2017}}

@article{Geneaux2021,
	author = {Romain G\'{e}neaux and Hung-Tzu Chang and Adam M. Schwartzberg and Hugo J. B. Marroux},
	doi = {10.1364/OE.412117},
	journal = {Opt. Express},
	month = {Jan},
	number = {2},
	pages = {951--960},
	publisher = {Optica Publishing Group},
	title = {Source noise suppression in attosecond transient absorption spectroscopy by edge-pixel referencing},
	url = {https://opg.optica.org/oe/abstract.cfm?URI=oe-29-2-951},
	volume = {29},
	year = {2021}}

@article{FerrayJPB1988,
	author = {M.~Ferray and A.~L'Huillier and X.F.~Li and L.A.~Lompre and G.~Mainfray and C.~Manus},
	journal = {Journal of Physics B: Atomic, Molecular and Optical Physics},
	pages = {L31-L35},
	title = {Multiple-harmonic conversion of 1064 nm radiation in rare gases},
	url = {https://iopscience.iop.org/article/10.1088/0953-4075/21/3/001/meta},
	volume = {\textbf{21}},
	year = {1988}}

@article{McPhersonJOSAB1987,
	author = {A.~McPherson and G.~Gibson and H.~Jara and U.~Johann and T.~S.~Luk and I.~A.~McIntyre and K.~Boyer and C.~K.~Rhodes},
	doi = {10.1364/JOSAB.4.000595},
	journal = {Journal of the Optical Society of America B},
	pages = {595-601},
	title = {Studies of multiphoton production of vacuum-ultraviolet radiation in the rare gases},
	url = {https://www.osapublishing.org/josab/fulltext.cfm?uri=josab-4-4-595},
	volume = {4},
	year = {1987}}

\clearpage
\onecolumngrid
\section{Supplemental Material}

\subsection{XUV absorption spectrum}

\begin{figure*}[htbp!]
    \centering
    \includegraphics[width=0.6\linewidth]{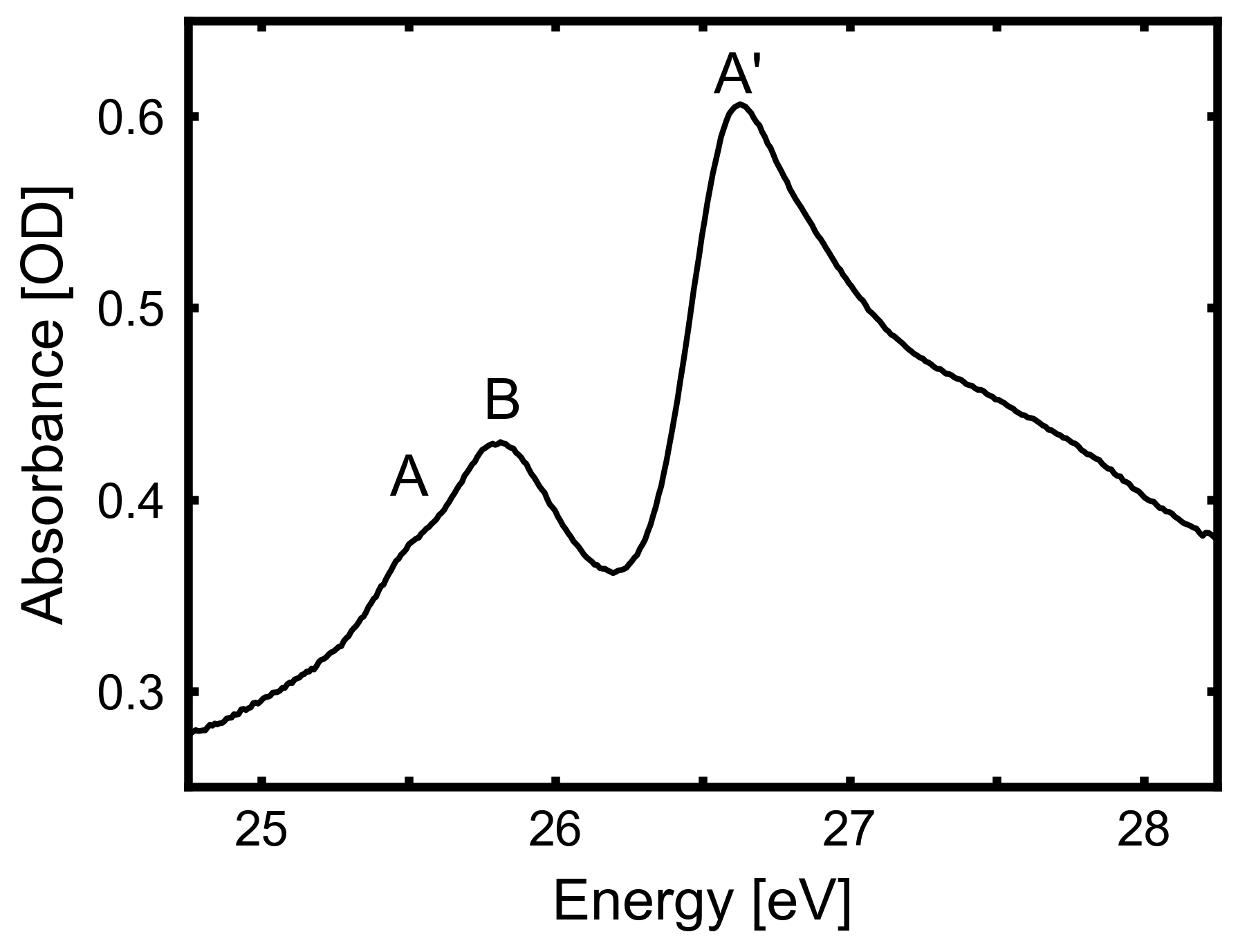}
    \caption{XUV absorption spectrum of the $\pi$-SnS thin film. Excitonic features labeled as in \cite{Taniguchi1990}.}
    \label{fig:static}
\end{figure*}

Figure~\ref{fig:static} shows the static XUV absorption spectrum of the $\pi$-SnS thin film. The spectrum displays a structured absorption profile originating from transitions from the 4$d$ core levels into the unoccupied conduction-band manifold and core excitonic states A, B and A$'$ \cite{Taniguchi1990}.

\subsection{Raman measurements}\label{raman}

Raman measurements were performed on the $\pi$-SnS sample, after all transient absorption were completed, in order to verify that there had been no change of the SnS phase. Raman spectra were collected using a Horiba LabRAM ARAMIS confocal microscope with 532 nm excitation, a ×100 objective lens, and an 1800 gr/mm grating.
\begin{figure*}[htbp!]
    \centering
    \includegraphics[width=0.6\linewidth]{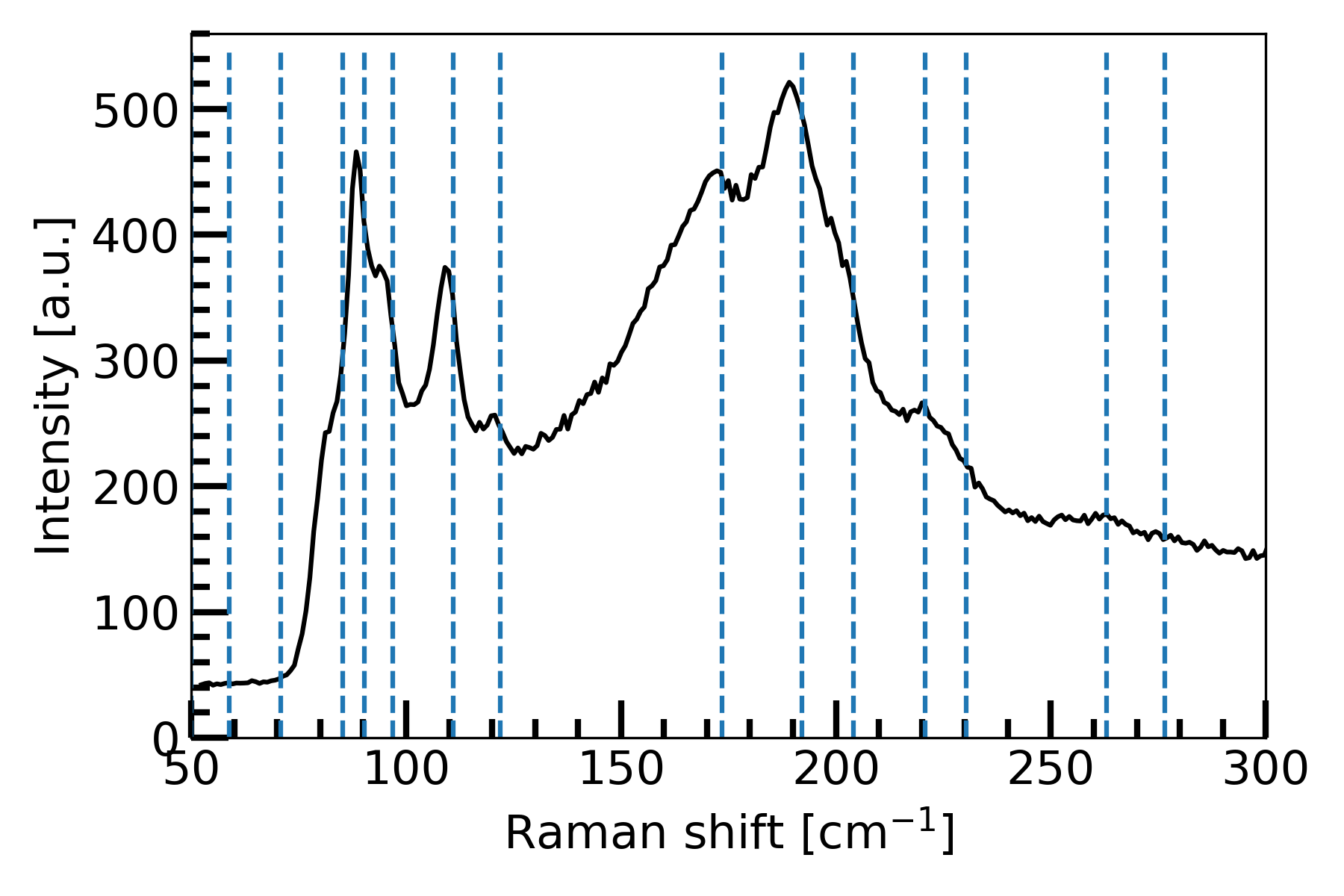}
    \caption{Raman spectrum of the $\pi$-SnS sample. Dashed lines denote Raman peaks for $\pi$-SnS as given in \cite{Guc20201}.
    }
    \label{fig:SI_SnSpi}
\end{figure*}

\begin{figure*}[htbp!]
    \centering
    \includegraphics[width=0.6\linewidth]{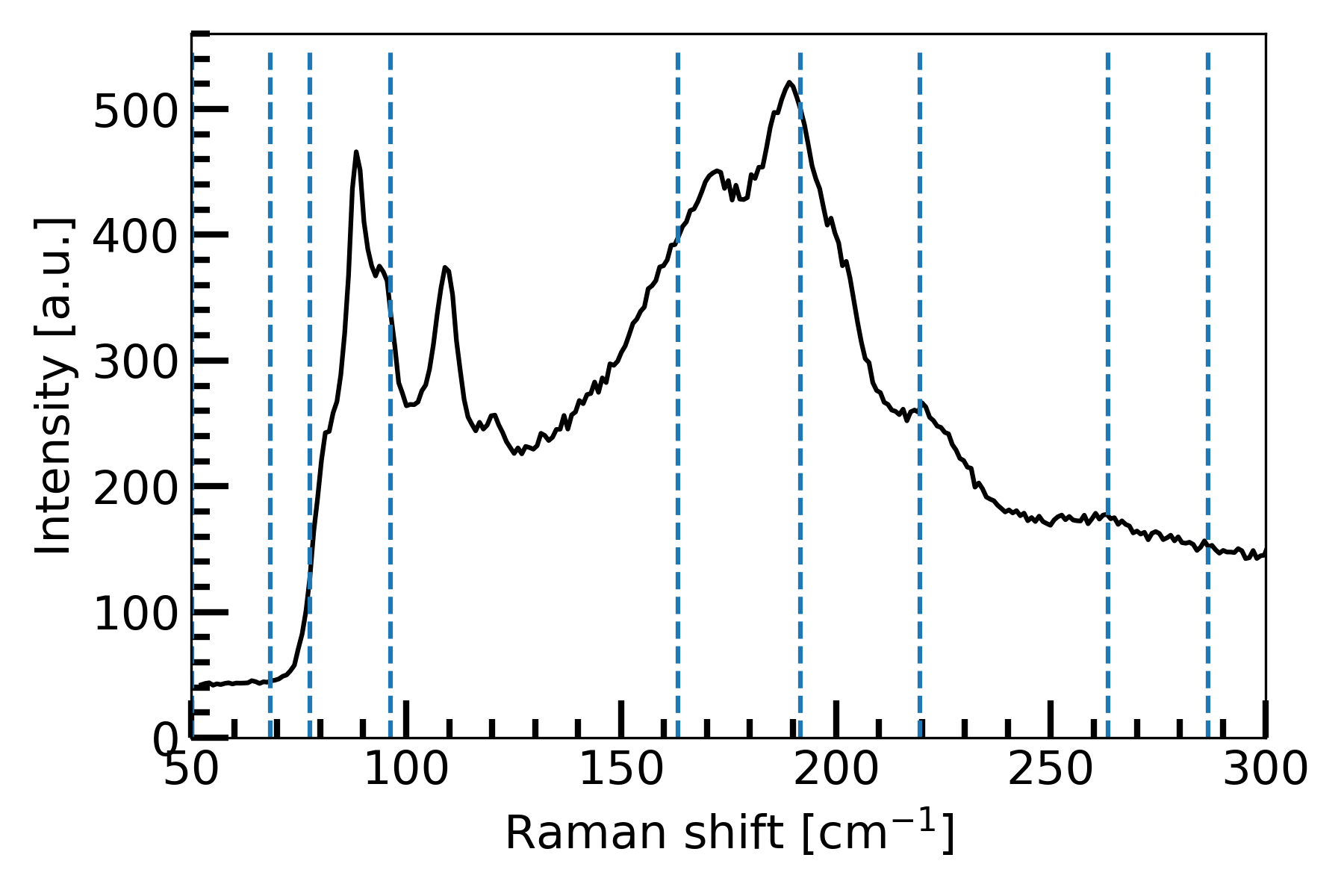}
    \caption{Raman spectrum of the $\pi$-SnS sample. Dashed lines denote Raman peaks for $\alpha$-SnS as given in \cite{Guc20201}.
    }
    \label{fig:SI_SnSalpha}
\end{figure*}

\subsection{X-ray diffraction characterization}\label{xrd}

X-ray diffraction measurements of SnS deposited on SiN were performed either with a Bruker AXS D8 Advance diffractometer with a Cu K$\alpha$ ($\lambda$K$\alpha$1 = 1.5406 Å, $\lambda$K$\alpha$2 = 1.54439 Å) radiation source, or with a Rigaku Miniflex 6G Benchtop Powder XRD with a Cu K$\alpha$ radiation source and HyPix-400MF Hybrid Pixel Array detector. For measurements acquired on the latter, the samples were mounted on an Al holder.

\begin{figure*}[htbp!]
    \centering
    \includegraphics[width=\linewidth]{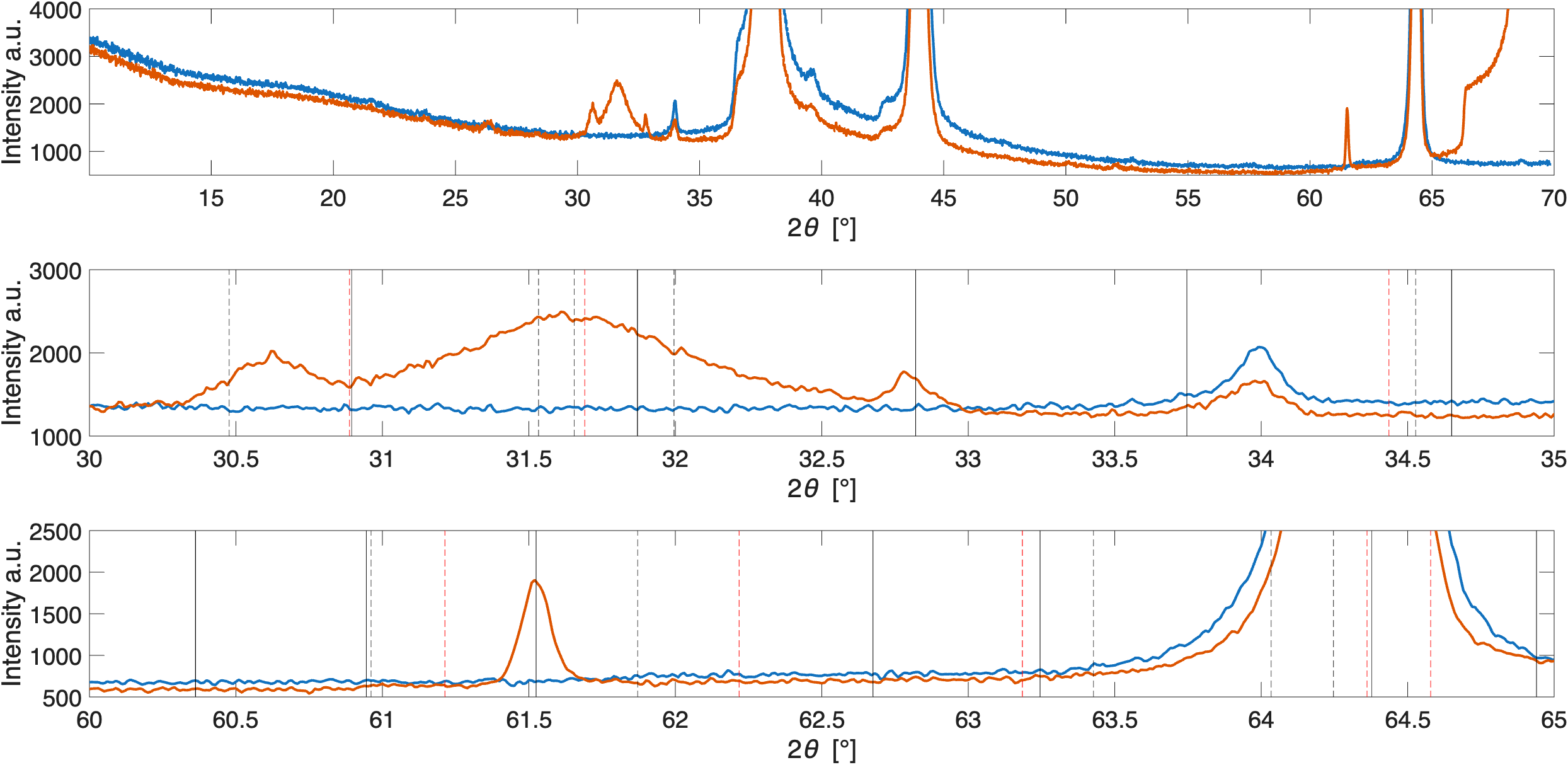}
    \caption{X-ray diffraction (XRD) patterns of 30 nm SnS thin films deposited on Si$_3$N$_4$ membranes.
The top panel shows the full $2\theta$ range, while the middle and bottom panels provide magnified views of the 30--35$^\circ$ and 60--65$^\circ$ regions, respectively.
Measured diffraction patterns of the SnS/SiN sample (orange) are shown together with reference measurements of the Al sample holder (blue), recorded under identical conditions. Vertical black solid, black dashed, and red dashed lines indicate calculated Bragg peak positions for $\pi$-SnS (cubic $P2_13$), $\alpha$-SnS (orthorhombic $Pnma$), and the Si$_3$N$_4$ substrate ($P31c$), respectively. The observed diffraction features are consistent with contributions from the $\pi$-SnS phase, superimposed on background scattering from the Al holder and the SiN membrane.}
    \label{fig:XRD_SnS}
\end{figure*}

Figure~\ref{fig:XRD_SnS} compares the diffraction pattern of the SnS on Si$_3$N$_4$ sample with that of the Al holder.
Several diffraction features in the measured spectra cannot be attributed solely to the holder or substrate and are consistent with Bragg reflections expected for SnS. To guide phase identification, calculated peak positions for cubic $\pi$-SnS ($P2_13$), orthorhombic $\alpha$-SnS ($Pnma$), and the SiN membrane ($P31c$) are indicated. The zoomed views highlight diffraction features at approximately $2\theta \approx 32.8^\circ$ and $61.5^\circ$, which coincide with prominent Bragg reflections expected for cubic $\pi$-SnS and are clearly observed in the sample spectrum. These reflections are not accounted for by the Al holder or the SiN membrane and are consistent with the presence of the $\pi$-SnS phase.

\subsection{Carrier cooling dynamics}
\label{sec:hot_phonon_model}

To describe the non-monotonic density dependence of the early-time carrier cooling rate, we employ a phenomenological two-temperature energy-balance model that captures the competition between electron--optical-phonon coupling and density-dependent carrier--carrier scattering processes.

Following ultrafast photoexcitation, carrier--carrier scattering rapidly establishes a thermalized electronic distribution characterized by an effective electron temperature $T_e(t)$. On the femtosecond timescale relevant here, the dominant channel for electronic energy relaxation is coupling to a subset of longitudinal optical (LO) phonons. These LO phonons act as an intermediate energy reservoir, they are efficiently populated by hot electrons but can only relax their energy to the rest of the lattice through comparatively slow anharmonic decay into lower-energy acoustic phonons \cite{Klemens1966}. We model this dynamics using coupled energy-balance equations for the electron and LO-phonon subsystems, expressed in terms of temperature deviations from the ambient lattice temperature $T_0$:
\begin{equation}
\Delta T_e(t) = T_e(t) - T_0, \qquad
\Delta T_{LO}(t) = T_{LO}(t) - T_0.
\end{equation}
The governing equations are,
\begin{align}
C_e(n)\frac{d\Delta T_e}{dt} &=
- G\bigl(\Delta T_e - \Delta T_{LO}\bigr)
- P_{cc}(n,\Delta T_e)\,,
\label{eq:Te_ode}
\\[4pt]
C_{LO}\frac{d\Delta T_{LO}}{dt} &=
+ G\bigl(\Delta T_e - \Delta T_{LO}\bigr)
- \frac{1}{\tau_{LA}}\,\Delta T_{LO}.
\label{eq:TLO_ode}
\end{align}
Here $G$ is the electron-phonon coupling constant, $C_e(n)$ the electronic heat capacity, $C_{LO}$ the LO-phonon heat capacity and $\tau_{LO}$ the effective lifetime governing the removal of energy from the LO-phonon subsystem into longitudinal acoustic (LA) phonons. The term $P_{cc}(n,\Delta T_e)$ represents an additional density-dependent electronic energy-relaxation channel due to Auger cooling. For the carrier densities explored here ($n \sim 10^{20}$--$10^{21}\,\mathrm{cm^{-3}}$), the electronic system is assumed to be degenerate on the timescale over which the early-time cooling rate is extracted. In this regime, the electronic heat capacity is given by,
\begin{equation}
C_e = \frac{\pi^2}{3}k_B^2 T_e D(E_F),
\end{equation}
where $D(E_F)$ is the density of states at the Fermi level. For a three-dimensional parabolic band, $D(E_F)\propto \sqrt{E_F}$ and $E_F\propto n^{2/3}$, giving,
\begin{equation}
C_e(n) \sim n^{1/3}.
\label{eq:Ce_scaling}
\end{equation}

\subsection{Carrier density calculation}

The average photoinduced charge carrier density $\Delta N$ was calculated as \cite{Cushing2018}:
\begin{equation}
    \Delta N = F\frac{\lambda}{hc}\frac{1-R}{l} \left( 1-e^{-\alpha l}\right) \left(1+0.02e^{-\alpha l} \right)
    \label{SM:eq:ccdens}
\end{equation}
where F denotes laser fluence (provided as a function of pump energy and pump area in Table \ref{SM:tbl:ccdens_fluences}), $R$ is the reflectivity of the sample, $hc/\lambda$ is the energy of the photons, $l$ is the sample thickness, and $\alpha$ is the estimated absorption coefficient for $\pi$-SnS at the relevant wavelength from spectroscopic ellipsometry. Back-reflections at the rear SnS-SiN interface are taken into account by the second exponential term, using Snell's law. The values used in Eq. \ref{SM:eq:ccdens} are given in Table \ref{SM:tbl:ccdens_params}. Pump energy-dependent fluences and $\Delta N$ are provided in Table \ref{SM:tbl:ccdens_fluences}.

The size of the pump beam at focus was measured using a camera as a function of pump energy, allowing a more accurate calculation of the laser fluence (see Table \ref{SM:tbl:ccdens_fluences}).

\begin{figure*}[htbp!]
    \centering
    \includegraphics[width=0.6\linewidth]{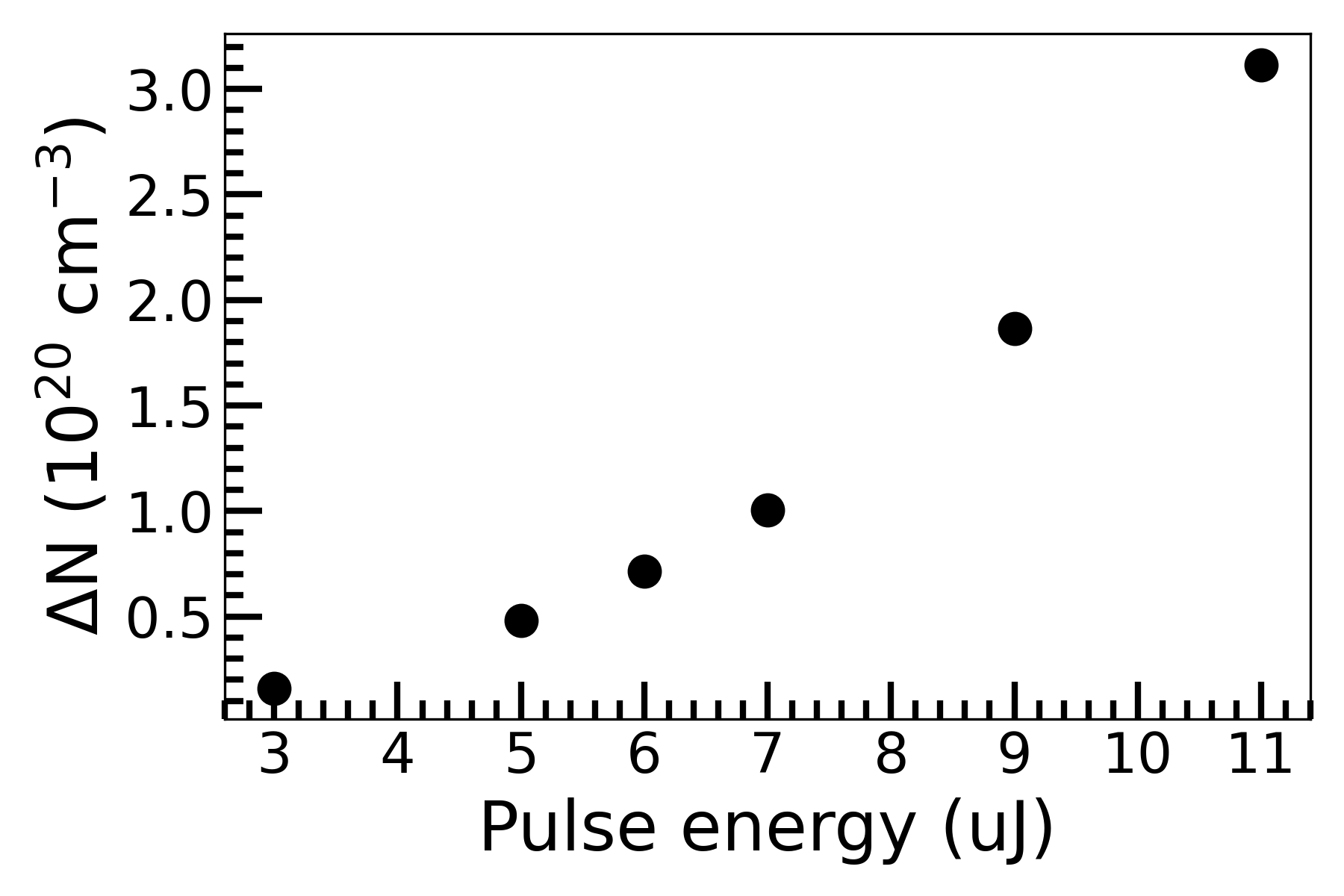}
    \caption{Calculated photoinduced charge carrier density as a function of pump pulse energy.}
    \label{fig:SM_ccDensPlot_2}
\end{figure*}

\begin{table}[h]
\begin{tabular}{|c|c|}
\hline
\textbf{Parameter} & \textbf{Value}  \\ \hline
$\lambda$ & 755 [nm]  \\ \hline
$R$ & 0.7 [-]  \\ \hline
$l$ & 30 [nm]  \\ \hline
$\alpha$ & $4\times10^{3}$ [cm$^{-1}$] \\ \hline
\end{tabular}
\caption{\textbf{Charge carrier density calculation parameters.} }
\label{SM:tbl:ccdens_params}
\end{table}

\begin{table}[h]
\begin{tabular}{|c|c|c|c|}
\hline
$E_P$ [$\mu$J] & $A$ [$10^{-4}$ cm$^2$] & $F$ [mJ/cm$^2$] & $\Delta N$ [$10^{20}$cm$^{-3}$] \\ \hline
3 & 8.63 & 3.48 & 0.161 \\ \hline
5 & 4.80 & 10.4 & 0.481 \\ \hline
6 & 3.89 & 15.4 & 0.713 \\ \hline
7 & 3.23 & 21.7 & 1.00 \\ \hline
9 & 2.23 & 40.4 & 1.87 \\ \hline
11 & 1.63 & 67.4 & 3.11 \\ \hline
\end{tabular}
\caption{\textbf{Pump parameters.} Pump energy and area, and resulting fluence and charge carrier density. }
\label{SM:tbl:ccdens_fluences}
\end{table}

\subsection{Direct forbidden type bandgap extrapolation}\label{bandgap}

\begin{figure*}[htbp!]
    \centering
    \includegraphics[width=0.6\linewidth]{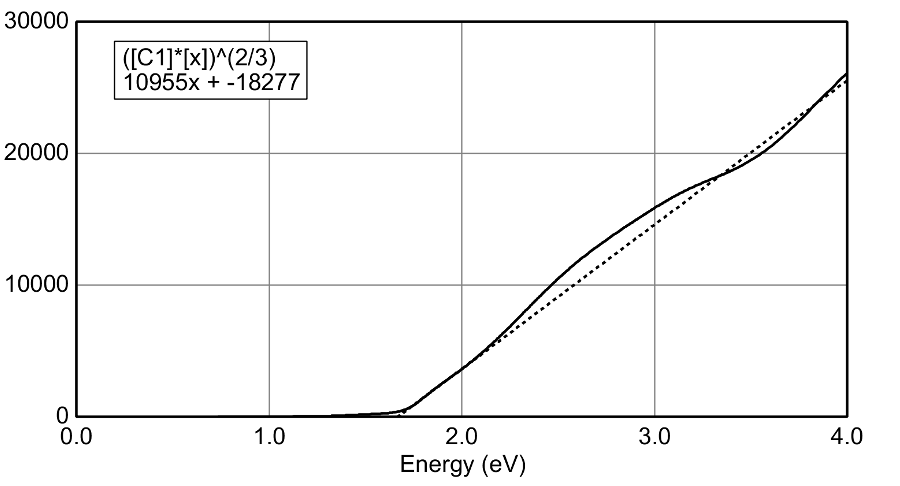}
    \caption{$(\alpha h \nu)^{2/3}$ vs $h \nu$ $\Rightarrow$ $E_g = 1.668$ eV.}
    \label{fig:ellips1}
\end{figure*}
A fit to a direct bandgap dependence is rather poor, in contrast:
\begin{figure*}[htbp!]
    \centering
    \includegraphics[width=0.6\linewidth]{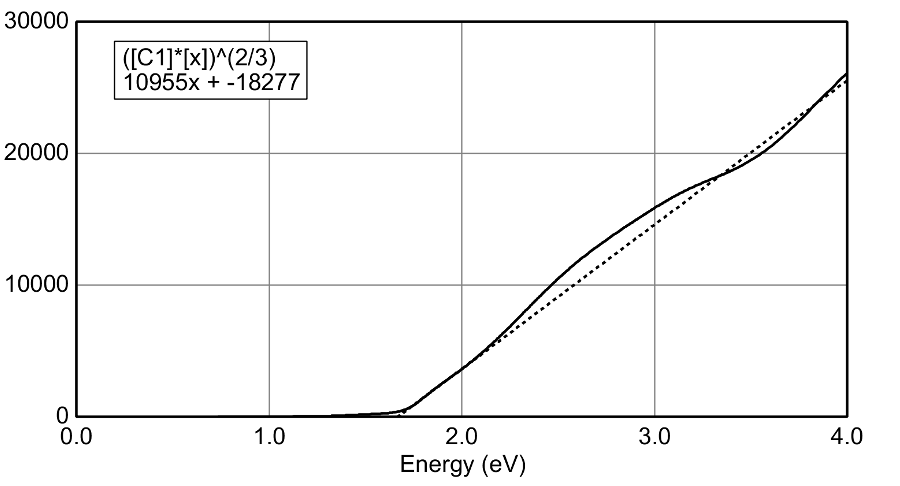}
    \caption{}
    \label{fig:ellips2}
\end{figure*}

\end{document}